\documentclass{elsart}
\usepackage{amssymb}
\usepackage{graphicx}
\usepackage[ps2pdf,colorlinks=true]{hyperref}

\usepackage{longtable}
\usepackage{multirow}
\usepackage{longtable}

\begin{document}
\begin{frontmatter}
\title{Intrinsic properties of high-spin band structures in triaxial nuclei}

\author{ S.~Jehangir$^{1,5}$, G.H.~Bhat$^{1,2,3}$, J.A.~Sheikh$^{1,3}$, R.~Palit$^{4}$ and P.A.~Ganai$^{5}$}
\address{$^1$ Department of Physics, University of Kashmir, Srinagar,
190 006, India \\
$^2$ Department of Physics, Govt. Degree College Kulgam, 192 231, India\\
$^3$ Cluster University Srinagar, Jammu and Kashmir, 190 001, India \\
$^4$  Department of Nuclear and Atomic Physics, Tata Institute of Fundamental Research, Colaba, Mumbai, 400 005,India\\
$^5$ Department of Physics, National Institute of Technology, Srinagar, 190 006, India
 }

\begin{abstract}
The band structures of $^{68,70}$Ge, $^{128,130,132,134}$Ce and
$^{132,134,136,138}$Nd are investigated using the triaxial projected
shell model (TPSM) approach. These nuclei depict forking of the ground-state
band into several s-bands and in some cases, both the lowest two observed s-bands depict
neutron or proton character. It was discussed in our earlier work that this anamoluos behaviour can be
explained by considering $\gamma$-bands based on two-quasiparticle
configurations. As the parent band and the $\gamma$-band built on it
have the same intrinsic structure, g-factors of the two bands are
expected to be similar. In the present work, we have undertaken a
detailed investigation of g-factors for the excited band structures of the
studied nuclei and the available data for a few high-spin states are shown to be
in fair agreement with the predicted values. 
\end{abstract}


\end{frontmatter}


\section{Introduction}

The wealth of high quality data obtained from high-spin nuclear spectroscopic
studies have provided invaluable information on the nature of the 
nuclear many-body system \cite{BM75,Mg49,fr01}. The rotation of triaxial nuclei is of topical interest because of 
the richness of the band structures plausible in these nuclei. The 
interplay of the rotational and the vibrational motions of the triaxial 
core with the gyromagnetic motion of the valance particles generates a wide
variety of nuclear excitation modes \cite{Ke01,g-Dy,Bonn96,je02,cl97,iw14}. A number of phenomena like signature splitting, gamma bands, forking of 
the ground band at intermediate spin, chiral rotation  and wobbling mode are 
related to the triaxial nuclear mode \cite{GH08,JG09,fr97,di00}. 

A major challenge in nuclear theory is to provide a microscopic and a unified
description of the collective and single-particle modes of excitations
in triaxial nuclei. Recently, it has been demonstrated that
microscopic approach of triaxial projected shell model (TPSM)
provides an excellent description of  various phenomena observed in
the triaxial nuclei \cite{SH99}.  For instance, the forking of the ground 
state band observed in some isotopes of Ge, Ce and Nd have
been explained by comparing the measured level schemes with calculated 
spectrum, obtained using the TPSM approach \cite{JG09,kum14}.
It was demonstrated that the anamolous behaviour of forking could
be explained by constructing quasiparticle band structures from a
triaxial mean-field potential. It has been shown that $\gamma$-bands are
built on each quasiparticle state as for the ground-state
configuration and the second observed two-quasiparticle aligned band is the
$\gamma$-band based on the two-quasiparticle aligned state in some nuclei
\cite{sugawara,pre123,pre2,pre3,RB79}. Since the
parent band and the $\gamma$-band built on it have the same
intrinsic structure, the observed two s-bands should have similar
properties. 

In order to probe the intrinsic properties of the aforementioned
excited structures, g-factors need to be evaluated that can 
be compared with the measured values. The purpose of the present 
work is to systematically evaluate the g-factors of the excited band
structures.
The present
work is organised in the following manner. In the next section, the
TPSM approach is briefly described for completeness and the emphasis
shall be on the evaluation of the electromagnetic properties. The results of the TPSM 
calculations are presented and discussed in section III and finally
the present work is summarised in section IV.

\section{Triaxial projected shell model approach}

TPSM approach has been demonstrated to correlate the high-spin 
properties of transitional nuclei quite well. The advantage of this model
is that computational requirements are quite minimal and a systematic
study of a large number of nuclei can be performed in a reasonable
time frame. The model employs the deformed intrinsic states of the
triaxial Nilsson potential as the basis configurations. These constitutes
an optimal basis set for a deformed system and a small subset of these
states is needed to have a satisfactory description of the near-yrast
spectroscopic properties. Since the deformed basis are defined in the intrinsic frame of reference
and don't have well defined angular-momentum, these basis are
projected onto states with well defined angular-momentum using the
angular-momentum projection technique \cite{ring80,HS79,HS80}. 
The three dimensional angular-momentum projection
operator is given by 
\begin{equation}
\hat P ^{I}_{MK}= \frac{2I+1}{8\pi^2}\int d\Omega\, D^{I}_{MK}
(\Omega)\,\hat R(\Omega),
\label{Anproj}
\end{equation}
with the rotation operator 
\begin{equation}
\hat R(\Omega)= e^{-i\alpha \hat J_z}e^{-i\beta \hat J_y}
e^{-i\gamma \hat J_z}.\label{rotop}
\end{equation}
 Here, $''\Omega''$ represents a set of Euler angles 
($\alpha, \gamma = [0,2\pi],\, \beta= [0, \pi]$) and the 
$\hat{J}^{'s}$ are angular-momentum operators.
The projected basis states
for the even-even system are composed of
 vacuum, two-proton, two-neutron and two-proton plus two-neutron
configurations, i.e.,
\begin{eqnarray}
\{ \hat P^I_{MK}\left|\Phi\right>, ~\hat P^I_{MK}~a^\dagger_{p_1}
a^\dagger_{p_2} \left|\Phi\right>, ~\hat P^I_{MK}~a^\dagger_{n_1}
a^\dagger_{n_2} \left|\Phi\right>,  \nonumber \\~\hat
P^I_{MK}~a^\dagger_{p_1} a^\dagger_{p_2} a^\dagger_{n_1}
a^\dagger_{n_2} \left|\Phi\right> \}, \label{basis}
\end{eqnarray}
where $\left|\Phi\right>$ in (\ref{basis}) represents the triaxial qp
vacuum state. In majority
of the nuclei, near-yrast spectroscopy, up to I=20 is well described
using the above basis space as one expects two-protons
to align after two-neutrons rather than four-neutrons considering the blocking argument.
However, we would like to add that this may not be the case for all the nuclei and there are
indications that four-neutron states may become important in the description of 
high-spin states in some rare-earth region nuclei \cite{156dy,156tdy}. 
TPSM calculations are performed in three stages. In the first stage,
triaxial basis are generated by solving the triaxially deformed Nilsson potential with
the deformation parameters of $\epsilon$ and $\epsilon'$. In the second stage, the
intrinsic basis are projected onto good angular-momentum states using the 
three-dimensional angular-momentum projection operator. In the third and final stage,
the projected basis are used to diagonalise the shell model Hamiltonian. 
The model
Hamiltonian consists of pairing and quadrupole-quadrupole interaction
terms \cite{KY95}, i.e.,

\begin{equation}
\hat H = \hat H_0 - {1 \over 2} \chi \sum_\mu \hat Q^\dagger_\mu
\hat Q^{}_\mu - G_M \hat P^\dagger \hat P - G_Q \sum_\mu \hat
P^\dagger_\mu\hat P^{}_\mu .
\label{hamham}
\end{equation}
The corresponding triaxial Nilsson Hamiltonian, which is used to generate the
triaxially-deformed mean-field basis  can be obtained by using
the Hartree-Fock-Bogoliubov (HFB) approximation, is given by
\begin{equation}
\hat H_N = \hat H_0 - {2 \over 3}\hbar\omega\left\{\epsilon\hat Q_0
+\epsilon'{{\hat Q_{+2}+\hat Q_{-2}}\over\sqrt{2}}\right\}.
\label{nilsson}
\end{equation}
In the above equation, $\hat H_0$ is the spherical single-particle
Nilsson Hamiltonian \cite{Ni69}.
 The monopole pairing strength $G_M$ 
is of the standard form
\begin{eqnarray}
G_M = {{(G_1 \mp G_2{{N-Z}\over A})}\frac{1}{ A}} (MeV),\label{pairing}
\end{eqnarray}
where the minus (plus) sign applies to neutrons (protons). 
In the present calculation, we choose  $G_1$ and $G_2$
such that the calculated gap parameters reproduce the experimental
mass  differences. The values $G_1$ and $G_2$, choosen in the present work, are 
$G_1 = 20.82$ and $G_2 = 13.58$ and are consistent with our earlier investigations  \cite{JG09,kum14,JS11,Zha15,PB02,GJ12,JY01,Ch12,YS08}.

 The Hamiltonian in Eq. (\ref{hamham}) is diagonalized using the projected 
basis of Eq. (\ref{basis}). 
The wave-functions obtained from the diagonalization are 
then used to evaluate the electromagnetic
transition probabilities.
The g-factor $ g(\sigma, I)$ is, 
generally, defined as
\begin{equation}
g(\sigma, I) = \frac {\mu(\sigma, I)}{\mu_N I} = g_\pi (\sigma,I) + g_\nu
(\sigma,I) , \label{ggfac}
\end{equation}
with $\mu(\sigma, I)$ being the magnetic moment of a state $(\sigma, I)$.
$g_\tau (\sigma, I), \tau = \pi$ or $\nu$, is given by
\begin{eqnarray}
g_\tau (\sigma, I) &=& {1\over{\mu_N I}} \langle \Psi^{\sigma}_{II} |
\hat \mu^\tau_z | \Psi^{\sigma}_{II} 
\rangle \nonumber \\
&=& {1\over{\mu_N \sqrt{I(I+1)}}} \langle \Psi^{\sigma}_{I} ||
\hat \mu^\tau || \Psi^{\sigma}_{I} 
\rangle \nonumber\\
&=& \frac{1}{\mu_N \sqrt{I(I+1)}} \left(
     g^{\tau}_l \langle \Psi^\sigma_I||\hat j^\tau ||\Psi^\sigma_I 
\rangle
     + (g^{\tau}_s - g^{\tau}_l) 
     \langle\Psi^\sigma_I||\hat s^\tau||\Psi^\sigma_I
\rangle \right) .
\end{eqnarray}
In our calculations, the following standard values for $g_l$ and
$g_s$ \cite{BM75} have been taken: $ g_l^\pi = 
1, g_l^\nu = 0, g_s^\pi = 5.586  $, and $ g_s^\nu =
-3.826 $ with an attenuation factor of $0.75$ for the spin components.
In the angular-momentum projection theory,
the reduced matrix element for $\hat m$ (with $\hat m$ being either 
$\hat j$ or $\hat s$) can be explicitly expressed as \cite{YJ02}
\begin{eqnarray}
&\langle& \Psi^{\sigma}_{I} || \hat m^\tau ||
\Psi^{\sigma}_{I} \rangle
\nonumber \\
&=& \sum_{K_i,K_f} f_{I K_i}^{\sigma} f_{I K_f}^{\sigma}
\sum_{M_i , M_f , M} (-)^{I - M_f}
\left(
\begin{array}{ccc}
I & 1 & I \\-M_f & M &M_i
\end{array} \right)
\langle \Phi | {\hat{P}^{I}}_{K_f M_f} \hat m_{1M}
\hat{P}^{I}_{K_i M_i} | \Phi \rangle\nonumber \\
 &=& (2I+1) \sum_{K_i,K_f} (-)^{I-K_f} f_{I K_i}^{\sigma} f_{I K_f}^{\sigma}
\nonumber \\
& & \times \sum_{M^\prime,M^{\prime\prime}}
\left(
\begin{array}{ccc}
I & 1 & I \\-K_{f} & M^\prime & M^{\prime\prime}
\end{array} \right)
\int d\Omega {\it D}_{M'' K_{i}} (\Omega)
\langle \Phi | \hat m_{1M'} \hat{R}(\Omega) | \Phi \rangle .
\end{eqnarray}
We finally obtain
\begin{eqnarray}
g_\tau (\sigma, I) &=& {1\over{\mu_N (I+1)}}
\sum_{K_i,K_f} f_{I K_i}^{\sigma} f_{I K_f}^{\sigma}
\sum_{M^\prime,M^{\prime\prime}} \langle I M^{\prime\prime} 1 M^\prime
 | I K_f \rangle \nonumber \\
&& \times \int d\Omega {\it D}_{M'' K_{i}} (\Omega)
\langle \Phi | \hat m_{1M'} \hat{R}(\Omega) | \Phi \rangle .
\end{eqnarray}

\begin{table}[htb]
\begin{center}
\caption{Axial and non-axial quadrupole deformation values, $\epsilon$
and $\epsilon'$ employed in the TPSM calculation for Ge, Ce and Nd nuclei. Axial 
deformations $\epsilon$ have been considered from \cite{Raman} for 
$^{68,70}$Ge nuclei, 
and from \cite{MN95} for $^{128,130,132,134}$Ce and $^{132,134,136,138}$Nd nuclei.  
The nonaxial values ($\epsilon'$) 
are chosen in such a way that band heads of the 
$\gamma$-bands are reproduced. The $\gamma$ deformation is related
to the above two parameters through $\gamma = \tan^{-1}(\epsilon'/\epsilon)$.} 
\vspace{0.9cm}
\begin{tabular}{|cccc|cccc|}
\hline
 A & $\epsilon$ & $\epsilon'$ & $\gamma$&A &$\epsilon$ & $\epsilon'$ & $\gamma$\\ \hline
 $^{68}$Ge & 0.220 & 0.160 & 36 & $^{70}$Ge & 0.235 & 0.145 & 31\\
  $^{128}$Ce & 0.250 & 0.120 & 26 & $^{132}$Nd & 0.267 & 0.120 & 24\\
 $^{130}$Ce & 0.225 & 0.120 & 28  & $^{134}$Nd & 0.200 & 0.120 & 31\\ 
  $^{132}$Ce & 0.183 & 0.100 & 29 & $^{136}$Nd & 0.158 & 0.110 & 35\\
  $^{134}$Ce & 0.150 & 0.100 & 34 & $^{138}$Nd & 0.170 & 0.110 & 33\\ \hline
\end{tabular}\label{defcd}
\end{center}
\end{table}
\section{Results and Discussion}
In the present work, a detailed TPSM study has been performed for
$^{68}$Ge, $^{70}$Ge, $^{128}$Ce, $^{130}$Ce, $^{132}$Ce, $^{134}$Ce,
$^{132}$Nd, $^{134}$Nd, $^{136}$Nd and $^{138}$Nd. TPSM basis
configurations are constructed by solving  triaxial
Nilsson potential with the deformation values depicted in Table 1. As
discussed in the previous section, since Nilsson potential is the
mean-field for quadrupole-quadrupole interaction, these deformation 
values also fix the interaction strength of the two-body interaction 
through self-consistency conditions \cite{KY95}. The deformation values
in Table 1 are same as used in our earlier investigations \cite{JG09,kum14}.  

\begin{figure}[htb]
 \centerline{\includegraphics[trim=0cm 0cm 0cm
0cm,width=0.8\textwidth,clip]{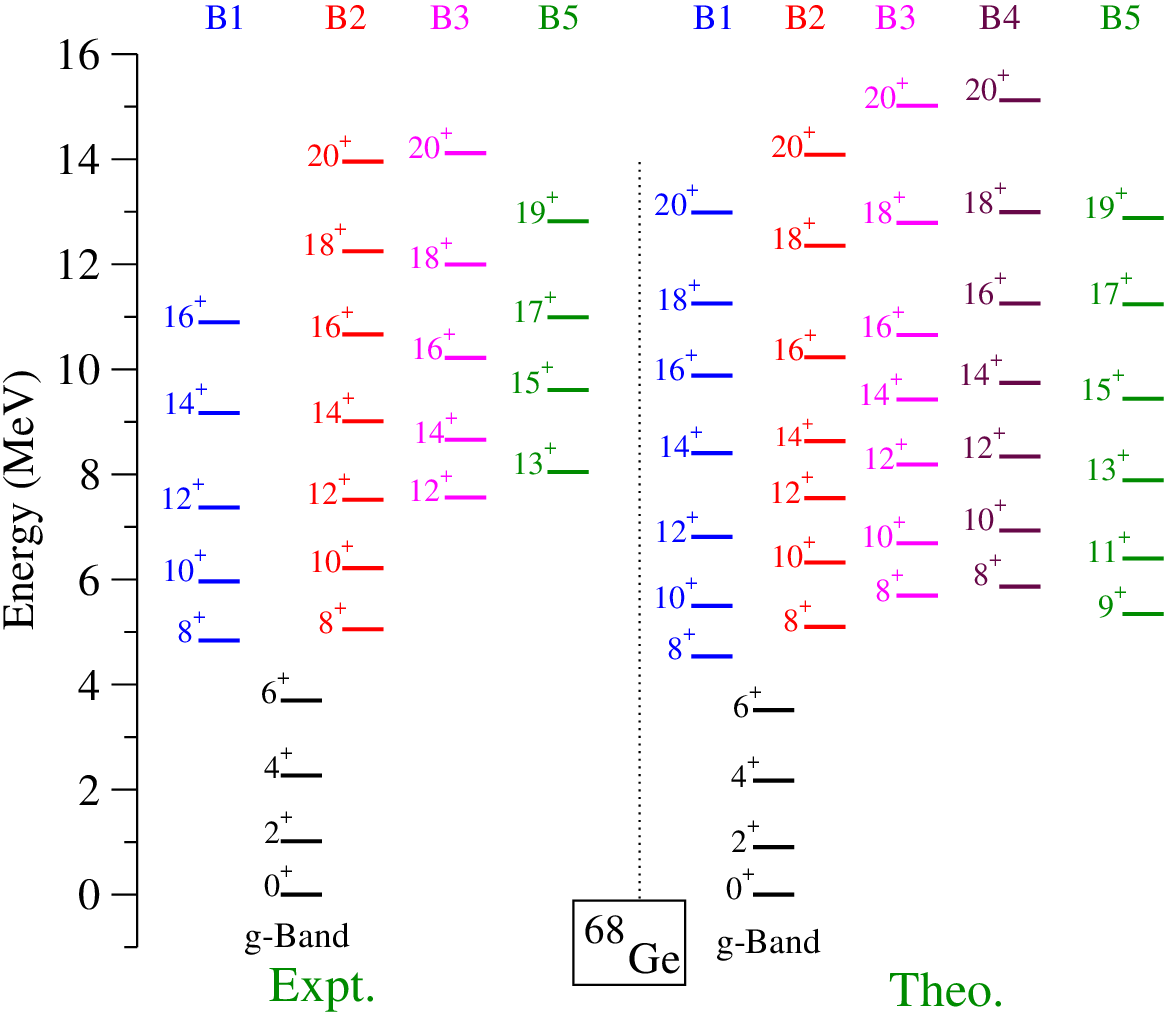}}
\caption{(Color online) Comparison of the calculated  energies with available experimental data for $^{68}$Ge. Data is taken from ref. \cite{68Ge}. }
\label{fig1}
\end{figure}

\begin{figure}[htb]
 \centerline{\includegraphics[trim=0cm 0cm 0cm
0cm,width=0.8\textwidth,clip]{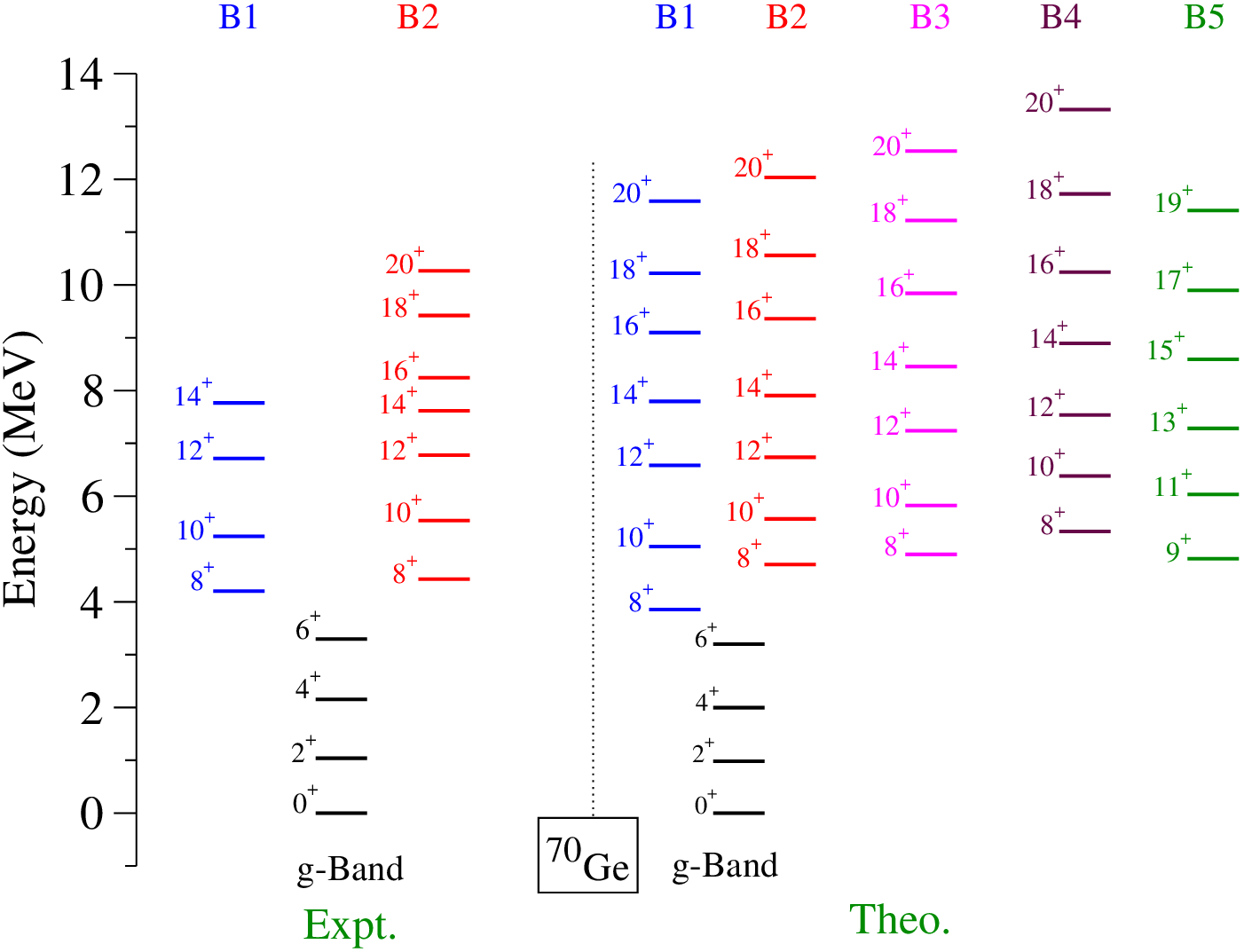}}
\caption{(Color online) Comparison of the calculated  energies with available experimental data for $^{70}$Ge. Data is taken from ref. \cite{kum14,DA70}.}
\label{fig1a}
\end{figure}

The obtained band structures of $^{68}$Ge and $^{70}$Ge, after
diagonalization of the shell model Hamitonian, Eq.~\ref{hamham} are 
depicted in Figs.~\ref{fig1}  and \ref{fig1a} along with the known experimental
data. It is noted in Fig.~\ref{fig1}  that the experimental ground-state band $^{68}$Ge forks into three
s-bands, labelled as B1, B2 and B3. In the earlier theoretical analysis using particle-rotor
model \cite{66Ge,68Ge}, it was predicted that the lowest s-band is the continuation
of the ground-state band and the other two s-bands are neutron and
proton aligned two-quasiparticle structures. However, experimental
study of relative g-factors using transient field technique indicated
that the bandheads (I=$8^+$) of the lowest two s-bands have both neutron structure \cite{66Ge,68Ge}. It was shown using
two-quasiparticle-plus-interacting boson model that this feature could
be explained by considering multipole interaction terms higher than
the quadrupole one \cite{IBM}.  In the TPSM results, shown on the right panel of
Fig.~\ref{fig1}, lowest band structures obtained after shell model
diagonalisation above spin, I=6 are plotted. It has been discussed in 
our earlier publication \cite{kum14} that band B1 is predominantly
composed of two-neutron aligned configuration with K=1 and band B2
is dominated by two-neutron aligned state, but with K=3. This
configuration with K=3 is the $\gamma$-band built on the parent 
two-neutron aligned state, having K=1. This interpretion shall be discussed
in detail later when presenting the results on g-factors.

\begin{figure}[htb]
 \centerline{\includegraphics[trim=0cm 0cm 0cm
0cm,width=0.8\textwidth,clip]{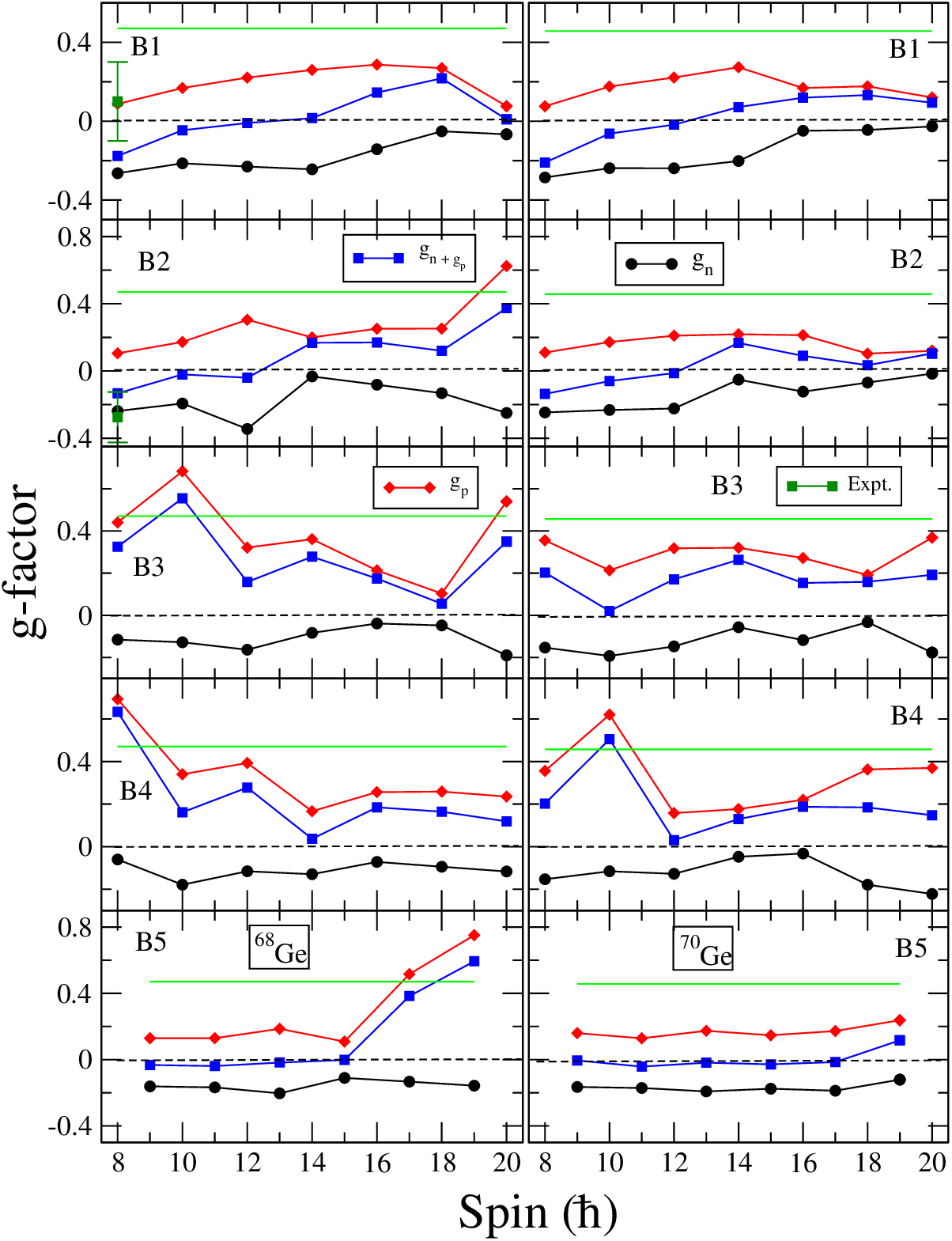}}
\caption{(Color online)  $^{68,70}$Ge g-factor with attenuated magnetic charges.  The green line shows $Z/A$.}
\label{fig2}
\end{figure}
\begin{figure}[htb]
 \centerline{\includegraphics[trim=0cm 0cm 0cm
0cm,width=0.8\textwidth,clip]{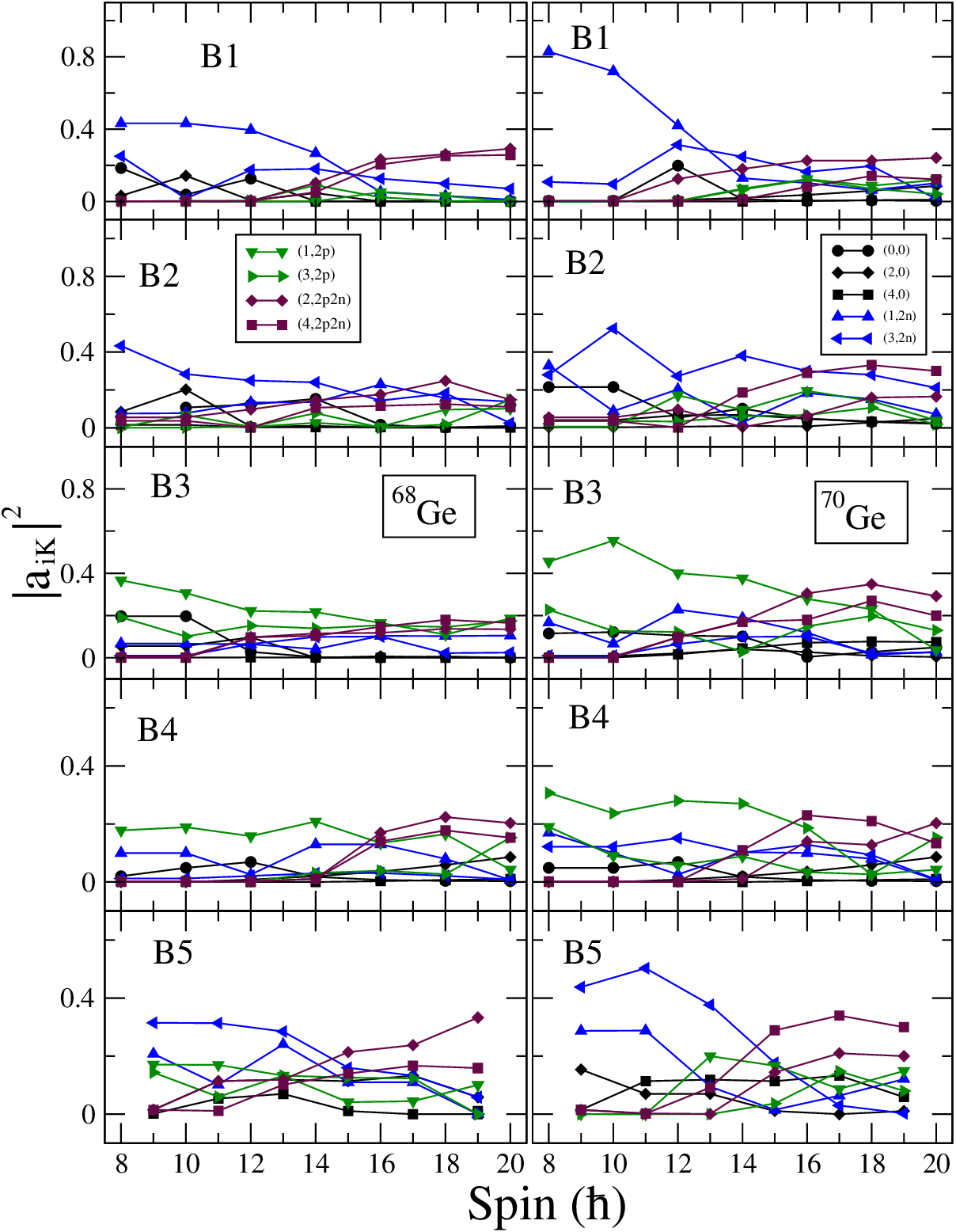}}
\caption{(Color online)   Probability of various projected K-configurations in the wavefunctions of the excited  bands in $^{68,70}$Ge. For clarity, only the lowest projected K-configurations in the wavefunctions of bands are shown and in the numerical calculations, projection has been performed from more than
forty intrinsic states.}
\label{fig3}
\end{figure}

The second s-band in $^{70}$Ge, labelled as band B2 in Fig.~\ref{fig1a}, was 
recently populated  \cite{kum14} and it has been shown using cranked shell
model (CSM)  approach that this band could not be attributed to the 
alignment of two-protons as crossing of this structure with the
ground-state band is expected around $\hbar \omega  =0.9 $MeV.
In the experimental data, both band B1 and B2 cross the ground-state
band at $\hbar \omega = 0.5$ MeV.  In CSM analysis, neutron crossing
occurs around this frequency and, therefore, both the bands are
expected to have neutron character. TPSM results indicate \cite{kum14} that band B2 is the $\gamma$-band based on two-neutron aligned structure and crosses the ground-state band at the same rotational frequency as that of the parent band B1. 

\begin{figure}[htb]
 \centerline{\includegraphics[trim=0cm 0cm 0cm
0cm,width=0.8\textwidth,clip]{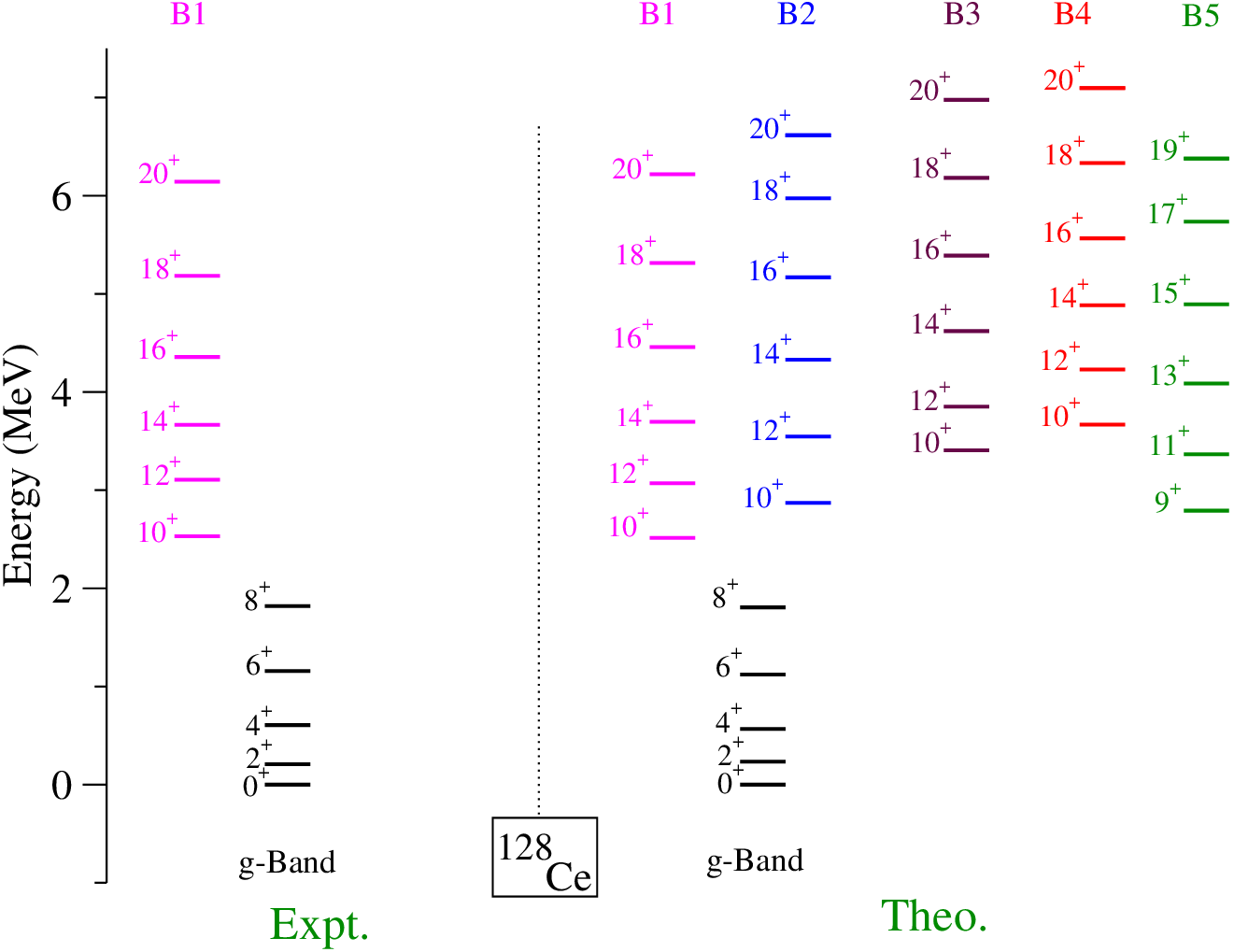}}
\caption{(Color online) Comparison of the calculated  energies with available experimental data for $^{128}$Ce. Data is taken from ref. \cite{Ep00}. }
\label{fig4}
\end{figure}
\begin{figure}[htb]
 \centerline{\includegraphics[trim=0cm 0cm 0cm
0cm,width=0.8\textwidth,clip]{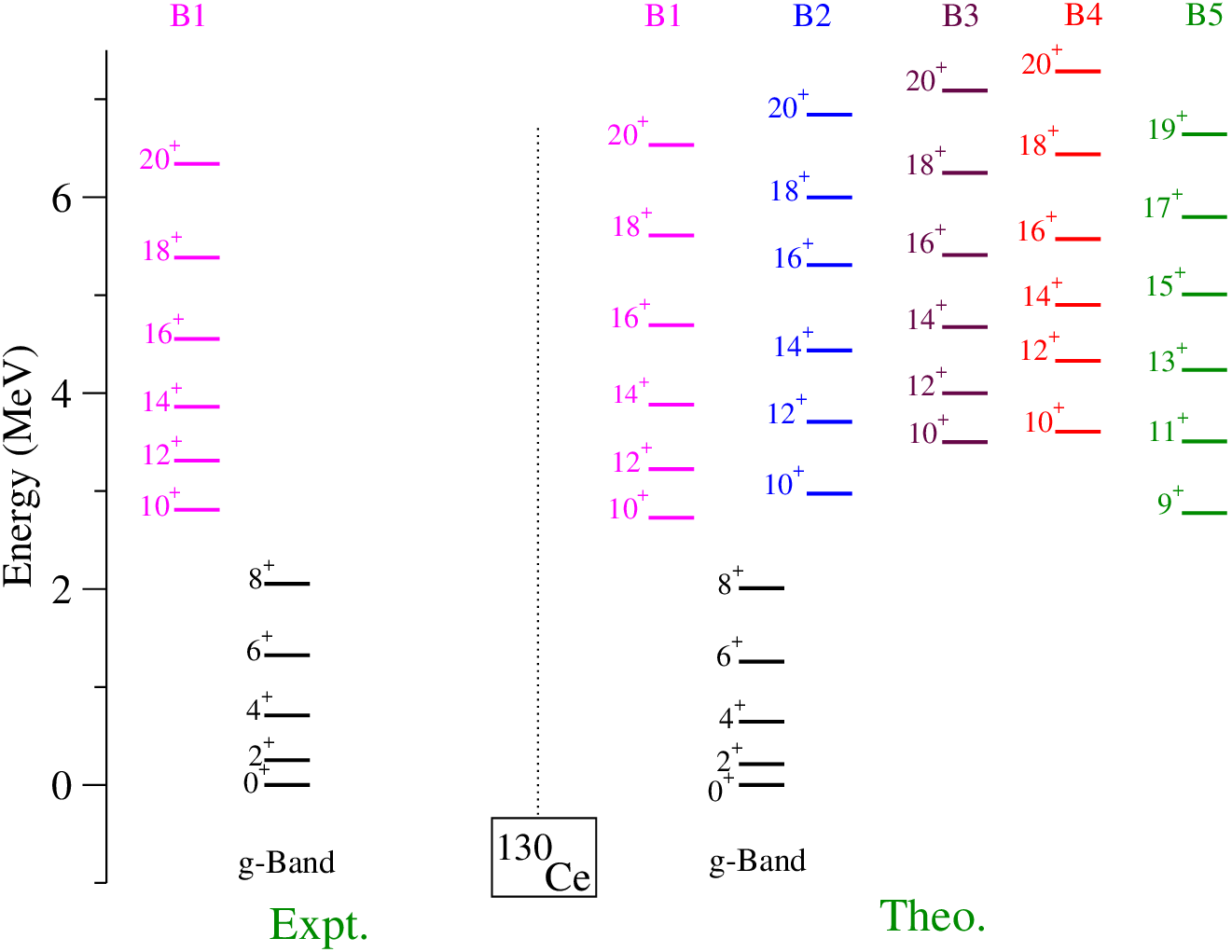}}
\caption{(Color online) Comparison of the calculated  energies with available experimental data for $^{130}$Ce. Data is taken from ref. \cite{Ce130}.}
\label{fig5}
\end{figure}
\begin{figure}[htb]
 \centerline{\includegraphics[trim=0cm 0cm 0cm
0cm,width=0.8\textwidth,clip]{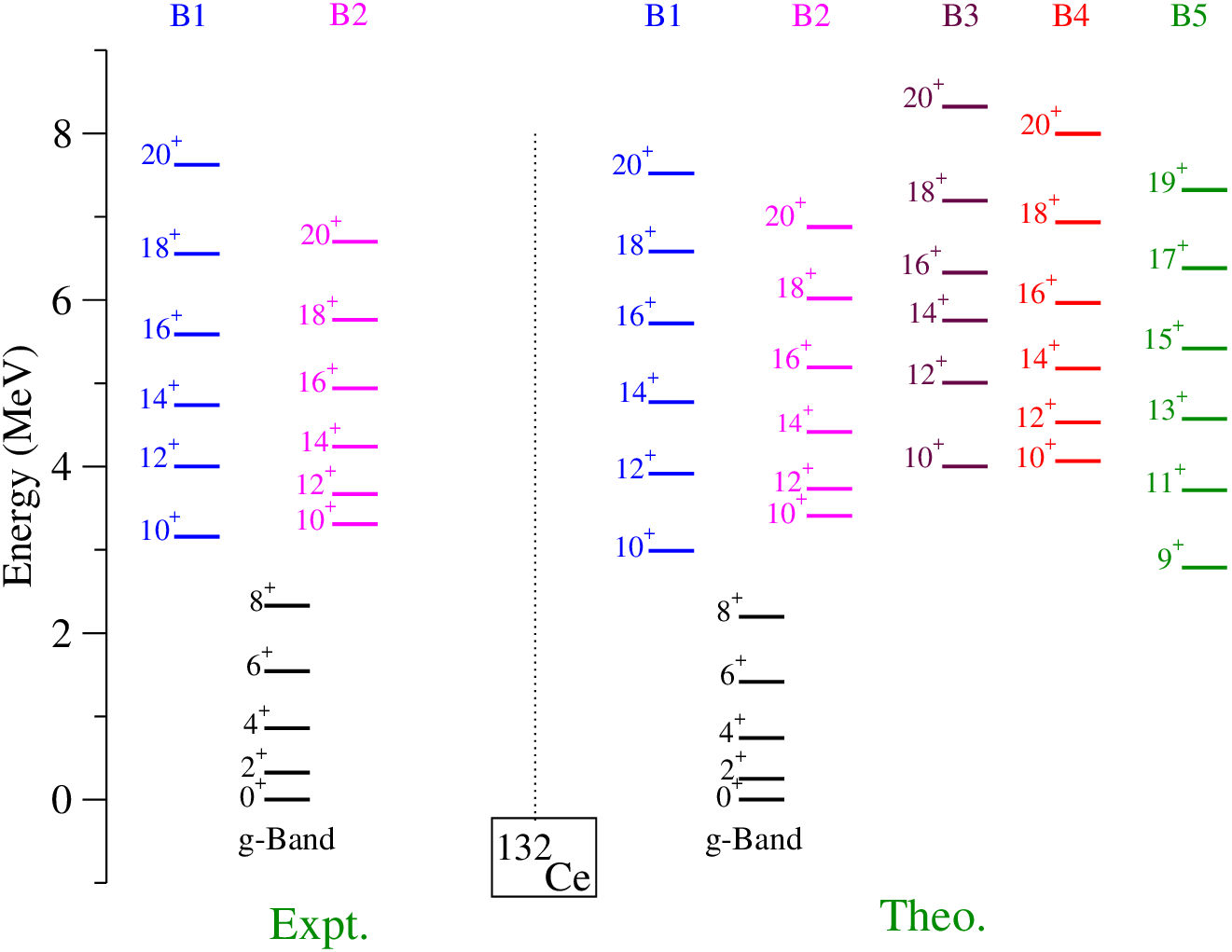}}
\caption{(Color online) Comparison of the calculated  energies with available experimental data for $^{132}$Ce. Data is taken from ref.  \cite{Pa05}.}
\label{fig6}
\end{figure}
\begin{figure}[htb]
 \centerline{\includegraphics[trim=0cm 0cm 0cm
0cm,width=0.8\textwidth,clip]{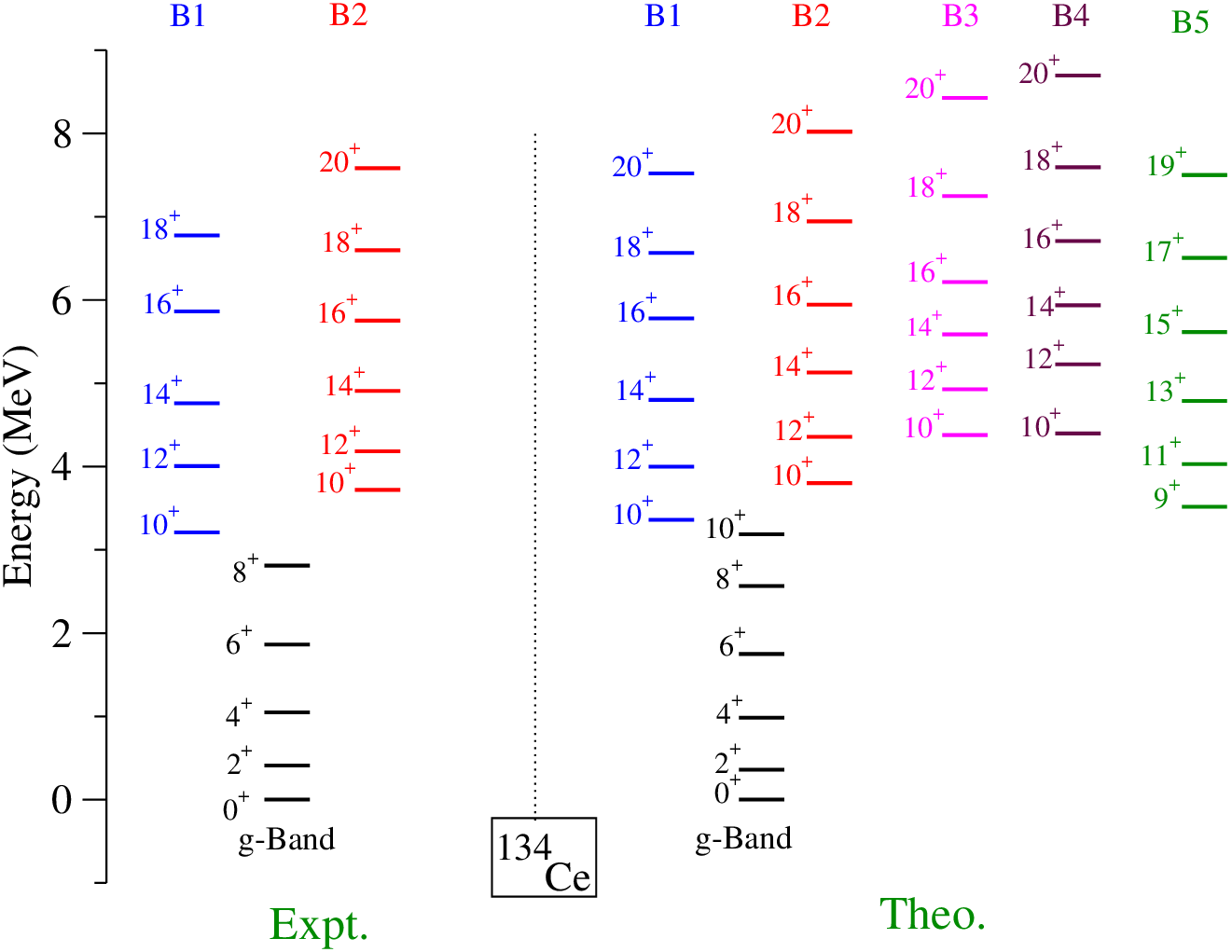}}
\caption{(Color online) Comparison of the calculated  energies with available experimental data for $^{134}$Ce. Data is taken from ref. \cite{Ce134}. }
\label{fig7}
\end{figure}

In order to quantify the intrinsic neutron and proton content, 
g-factors have been evaluated for the excited  band 
structures. As the single particle neutron and
proton gyromagnetic ratios have opposite signs with $g_n=-1.91$
and $g_p=2.79$, the measurement of g-factors provides information 
on the proton/neutron structure of a given state. The calculated g-factors,
using the TPSM wavefunction and the expression, Eq. \ref{ggfac},
for $^{68}$Ge and $^{70}$Ge are displayed in Fig.~\ref{fig2} for all the
excited band structures of Figs.~\ref{fig1} and \ref{fig1a}. 
Apart from the total g-factor, individual neutron and proton g-factor, labelled
as $g_n$ and $g_p$, and the rigid value of Z/A are also depicted in these figures.  
First of all, it is evident
from the figure that aside from some minor variation, the overall behaviour
of the g-factors with spin for the two nuclei is very similar. For
band B1 in both nuclei, the total g-factor for spin values up to I=12 is negative,
indicating that this band is dominated by neutron configuration. This
is evident from the wavefunction decomposition of the band structures
of two nuclei plotted in Fig.~\ref{fig3}.  Band B1 is noted to be
dominated by $(1,2n)$, which is a two-neutron aligned configuration
with K=1 up to I=12. For I=14 and above, it is observed that
$(2,2p2n)$ and $(4,2p2n)$ configurations, which are
four-quasiparticle states with K=2 and 4, respectively,
dominate. Since one-particle proton g-factor is much larger than the 
corresponding neutron g-factor, the total g-factor is positive for the
two-proton plus two-neutron state above I=14 in Fig.~\ref{fig3}. 
\begin{figure}[htb]
 \centerline{\includegraphics[trim=0cm 0cm 0cm
0cm,width=0.8\textwidth,clip]{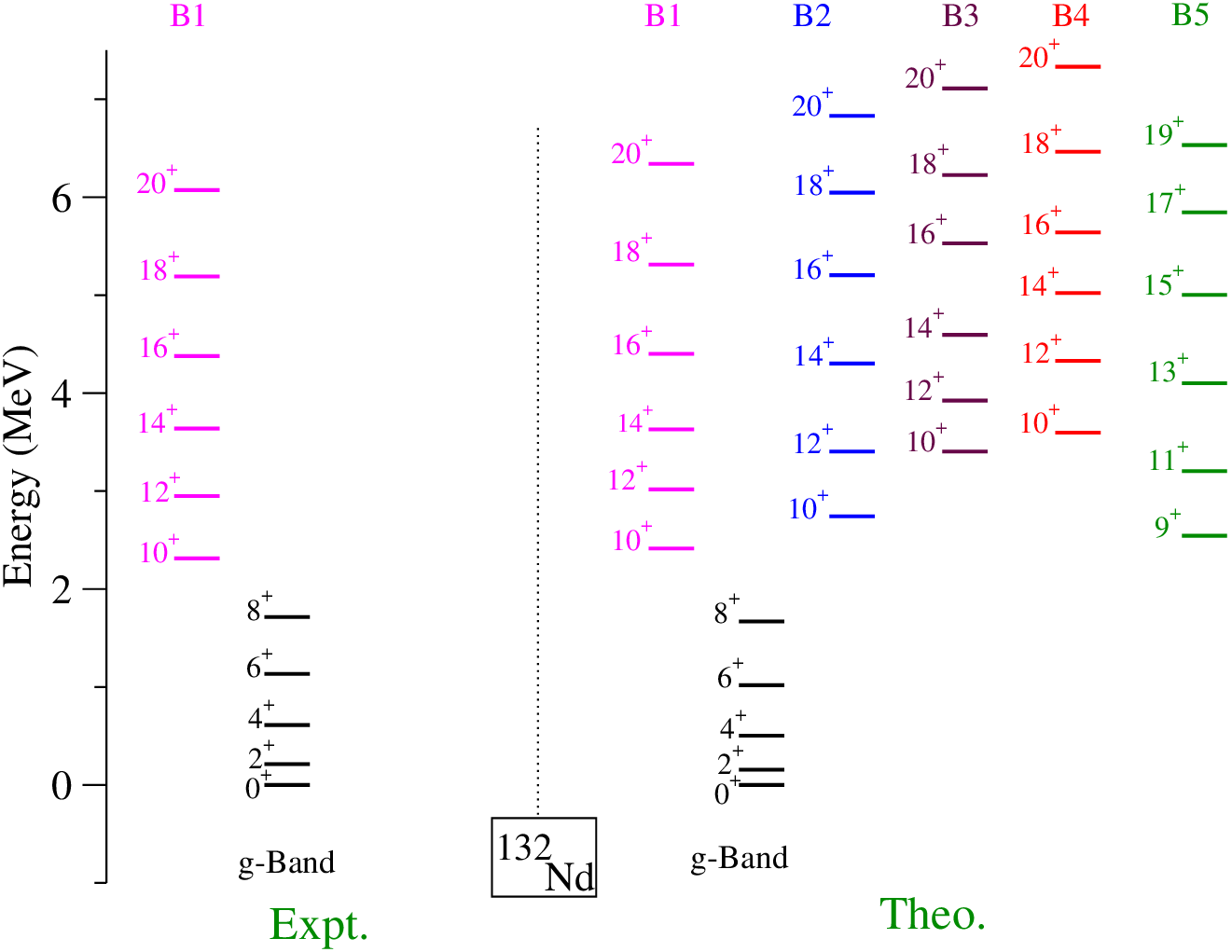}}
\caption{(Color online) Comparison of the calculated  energies with available experimental data for $^{132}$Nd. Data is taken from ref. \cite{Nd132}.}
\label{fig4a}
\end{figure}
\begin{figure}[htb]
 \centerline{\includegraphics[trim=0cm 0cm 0cm
0cm,width=0.8\textwidth,clip]{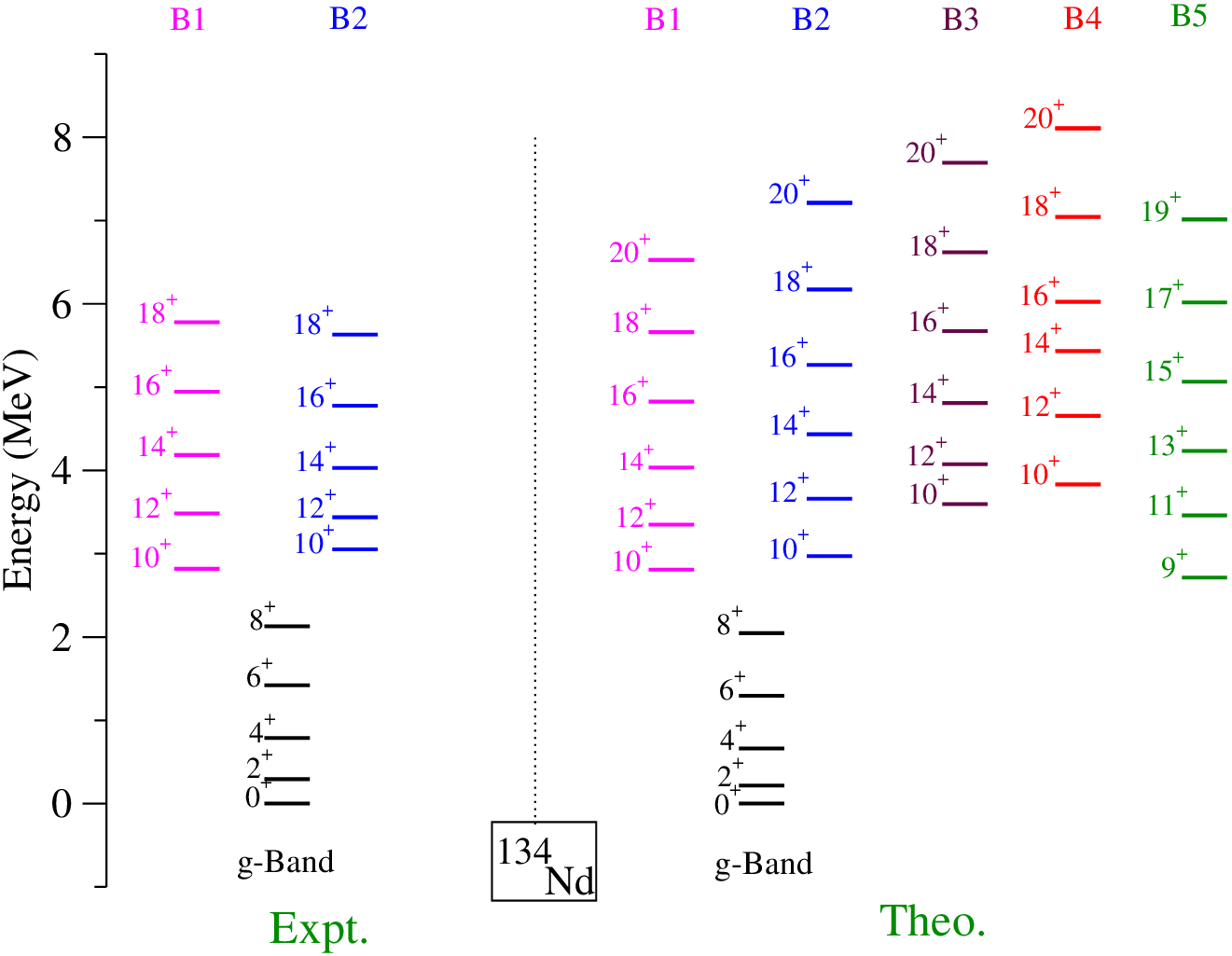}}
\caption{(Color online) Comparison of the calculated  energies with available experimental data for $^{134}$Nd. Data is taken from ref. \cite{Cd96}. }
\label{fig5a}
\end{figure}
\begin{figure}[htb]
 \centerline{\includegraphics[trim=0cm 0cm 0cm
0cm,width=0.8\textwidth,clip]{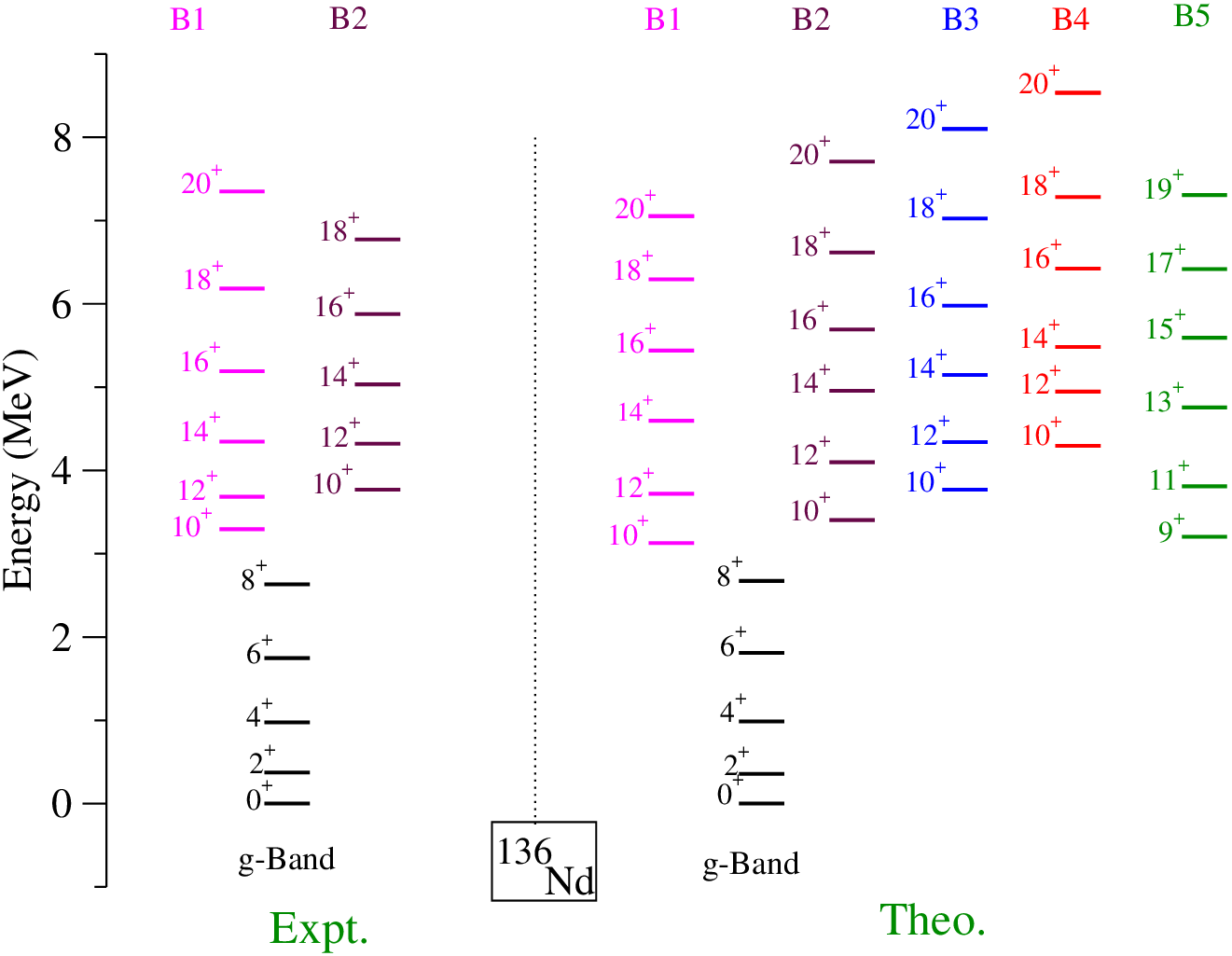}}
\caption{(Color online) Comparison of the calculated  energies with available experimental data for $^{136}$Nd. Data is taken from ref. \cite{Od02}.}
\label{fig6a}
\end{figure}
\begin{figure}[htb]
 \centerline{\includegraphics[trim=0cm 0cm 0cm
0cm,width=0.8\textwidth,clip]{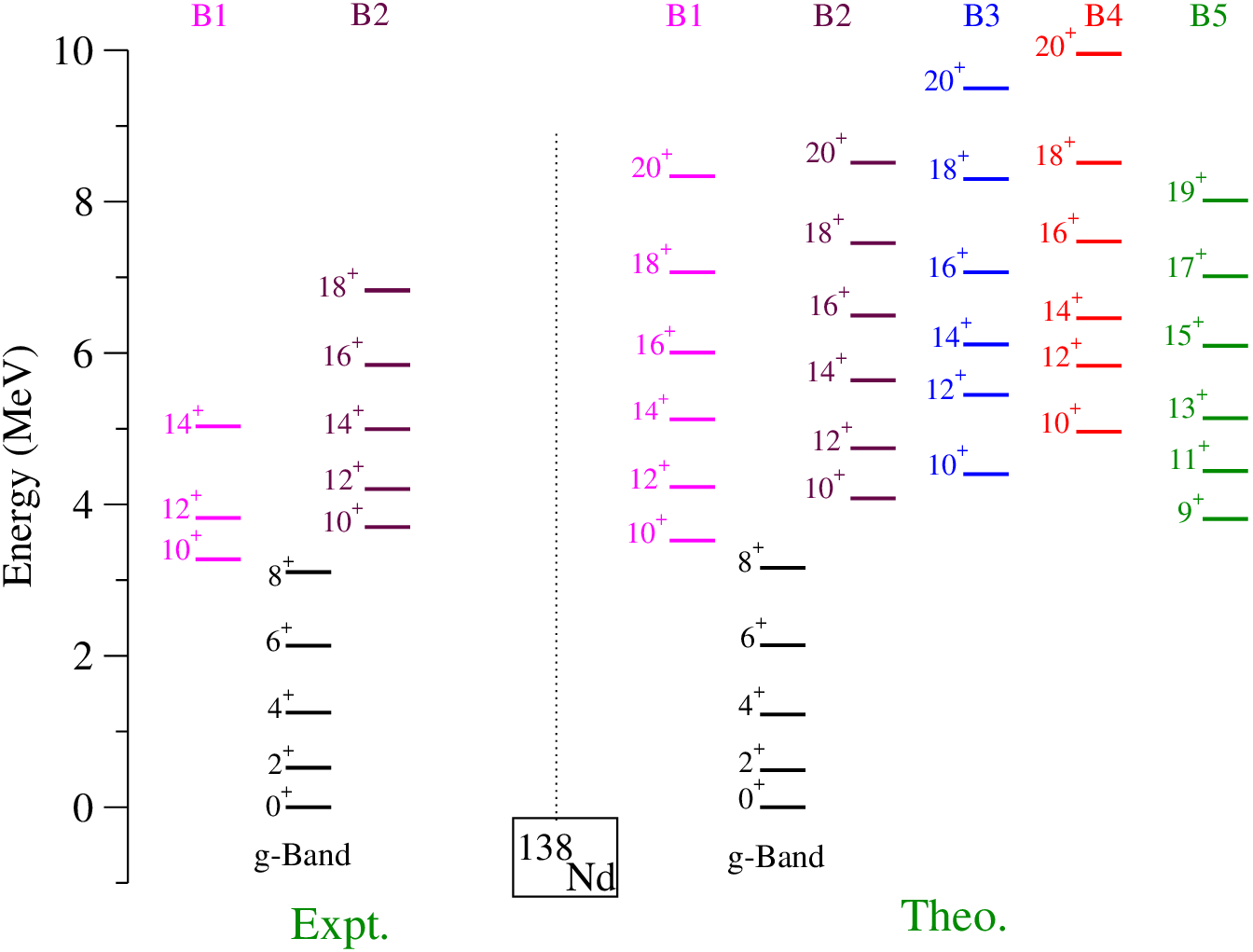}}
\caption{(Color online) Comparison of the calculated  energies with available experimental data for $^{138}$Nd. Data is taken from ref. \cite{Es02}. }
\label{fig7a}
\end{figure}

It is interesting to note in Fig.~\ref{fig2} that band B2 has a similar
g-factor values as that of  the band B1. The reason for this similarity
can be easily inferred from the wavefunction decomposition of
band B2 plotted in Fig.~\ref{fig3}. This band is dominated by $(3,2n)$ configuration,
which is the $\gamma$-band based on the two-neutron aligned
configuration, $(1,2n)$. Since the two configurations have similar
intrinsic structure, the g-factors of the two bands are expected to be 
similar. This naturally explains why the observed gyromagnetic ratios of the I=8 states
of bands B1 and B2 in $^{68}$Ge have similar g-factors. In the earlier work, higher
multipole interaction in the particle-rotor model picture was invoked \cite{IBM}
to explain this apparent anomally. It is not clear, at this stage, how this can be
related to the results obtained in the present work. The experimental g-factors for I=8 states of the bands B1 and B2 are also plotted in Fig.~\ref{fig2} for $^{68}$Ge. For band B1, the measured value is about g=0.1 \cite{IBM}, which indicates that the state has neutron character  and the TPSM calculations predict neutron character for the state with g = -0.26. For band B2, both measured and the calculated g-factors are negative and have similar values.
\begin{figure}[htb]
 \centerline{\includegraphics[trim=0cm 0cm 0cm
0cm,width=0.9\textwidth,clip]{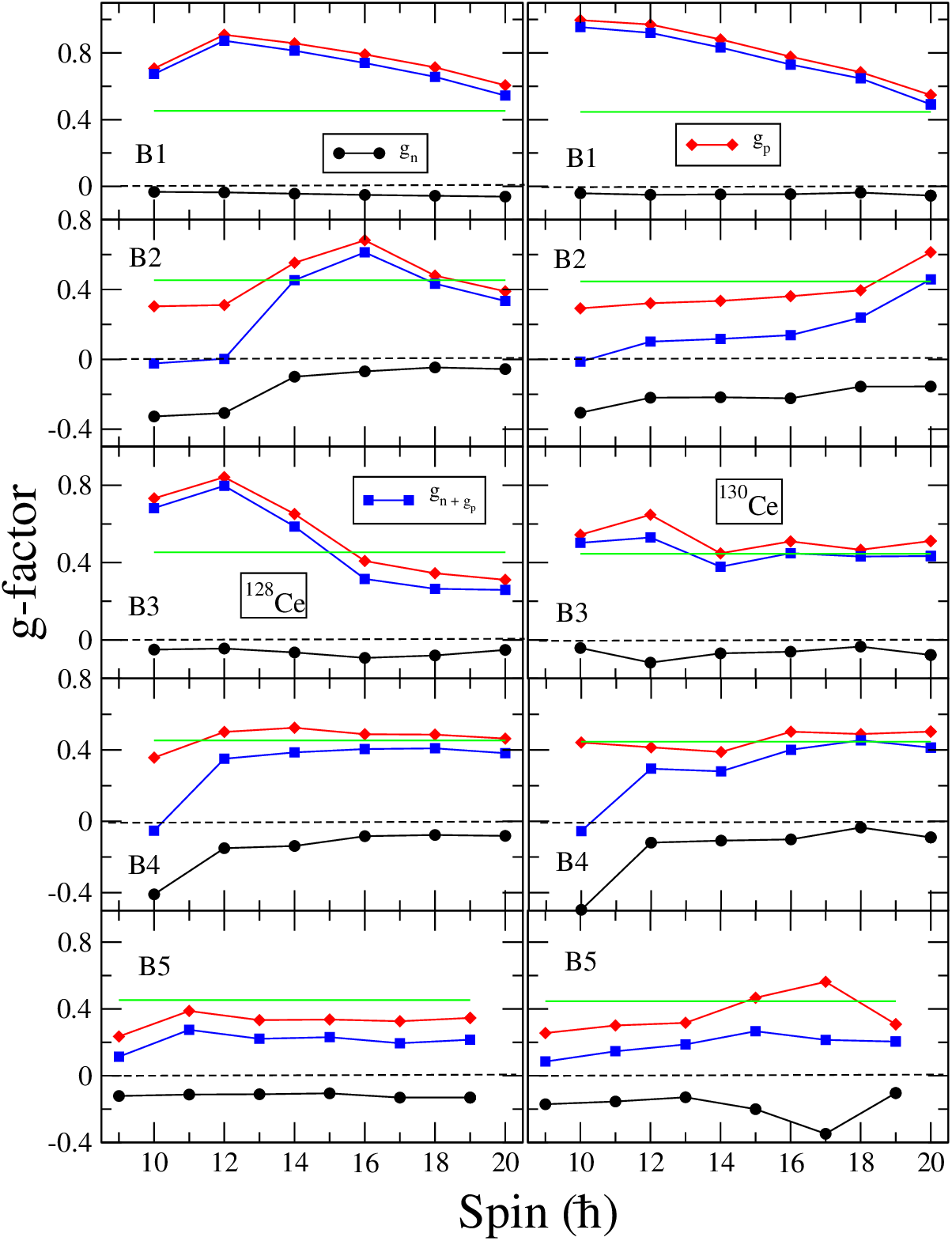}}
\caption{(Color online)  $^{128,130}$Ce g-factor with attenuated magnetic charges.  The green line shows $Z/A$.}
\label{fig8}
\end{figure}
\begin{figure}[htb]
 \centerline{\includegraphics[trim=0cm 0cm 0cm
0cm,width=0.9\textwidth,clip]{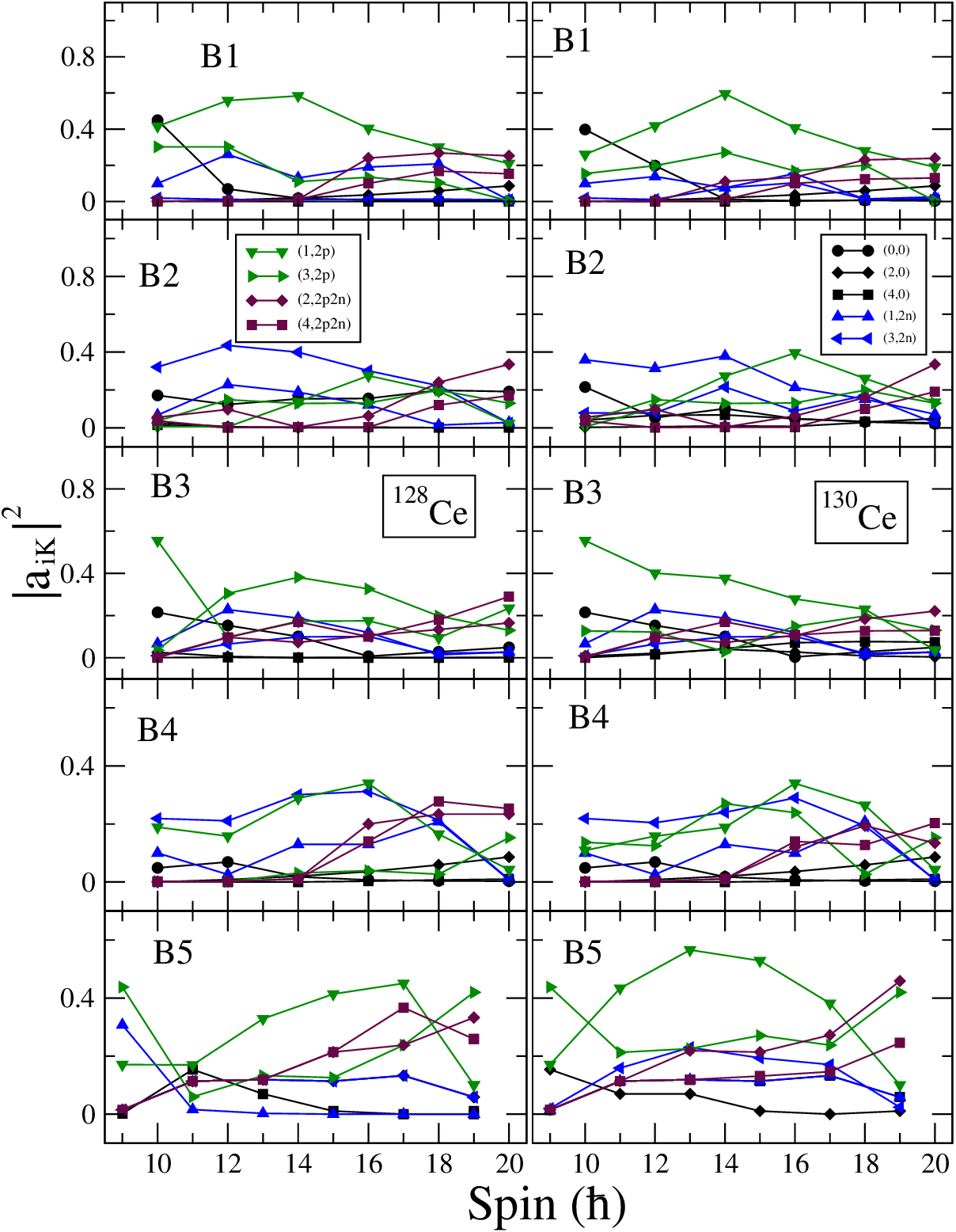}}
\caption{(Color online)   Probability of various projected K-configurations in the wavefunctions of the observed  bands for $^{128,130}$Ce. For clarity, only the lowest projected K-configurations in the wavefunctions of bands are shown and in the numerical calculations, projection has been performed from more than  
forty intrinsic states.}
\label{fig9}
\end{figure}
\begin{figure}[htb]
 \centerline{\includegraphics[trim=0cm 0cm 0cm
0cm,width=0.9\textwidth,clip]{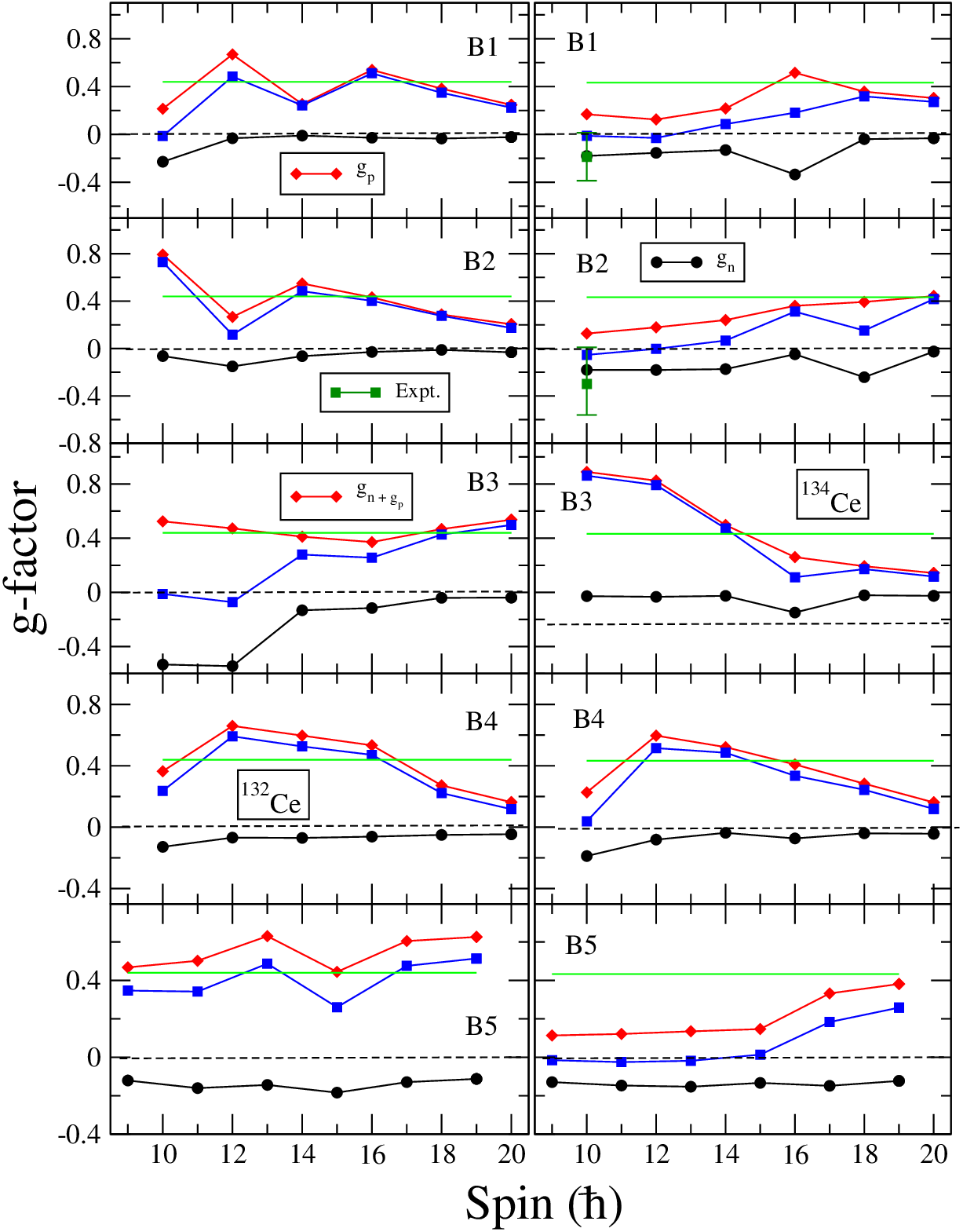}}
\caption{(Color online)  $^{132,134}$Ce g-factor with attenuated magnetic charges.  The green line shows $Z/A$.}
\label{fig10}
\end{figure}
\begin{figure}[htb]
 \centerline{\includegraphics[trim=0cm 0cm 0cm
0cm,width=0.9\textwidth,clip]{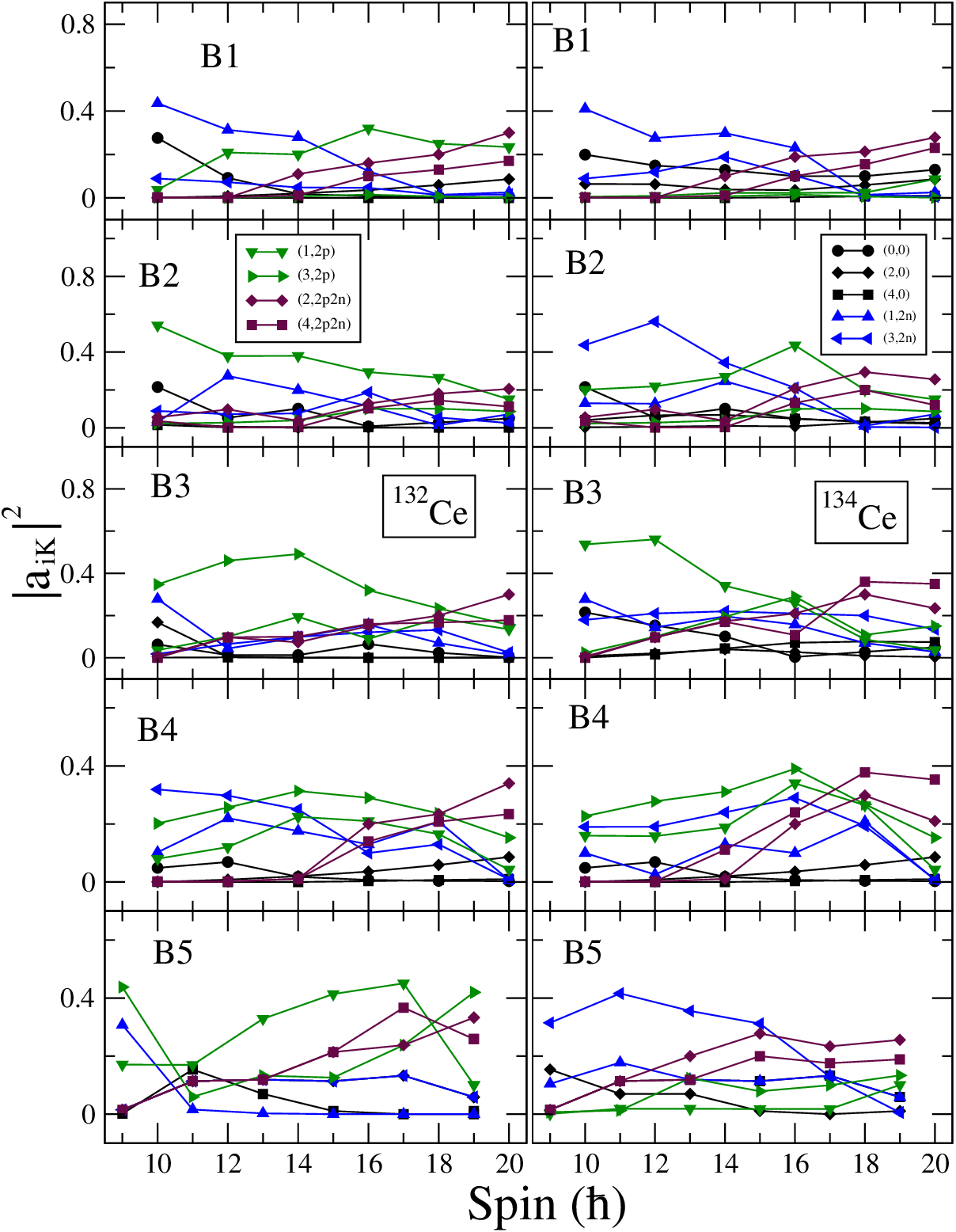}}
\caption{(Color online)   Probability of various projected K-configurations in the wavefunctions of the observed  bands for $^{132,134}$Ce. For clarity, only the lowest projected K-configurations in the wavefunctions of bands are shown and in the numerical calculations, projection has been performed from more than
forty intrinsic states.}
\label{fig11}
\end{figure}

The total g-factors of Bands B3 and B4 in Fig.~\ref{fig2} for both nuclei
are positive, indicating that proton component is dominant for these
band structures. This is evident from the wavefunction plot of
Fig.~\ref{fig3} with band B3 dominated by $(1,2p)$ component, which 
is a proton aligned configuration with K=1. The g-factors of band B4
are similar as that of band B3 and from the wavefunction
analysis, it is noted that band B4 is dominated by $(3,2p)$,
 which is a $\gamma$-band built on $(1,2p)$ aligned state. 
Band B5 for both the nuclei has a larger component of $(3,2n)$
configuration, but there are
also significant other contributions. This mixing of various
components leads to almost vanishing of  the  total g-factor for this
particular band structure, except for high spin states from I=17 to 19 in $^{68}$Ge.
\begin{figure}[htb]
 \centerline{\includegraphics[trim=0cm 0cm 0cm
0cm,width=0.9\textwidth,clip]{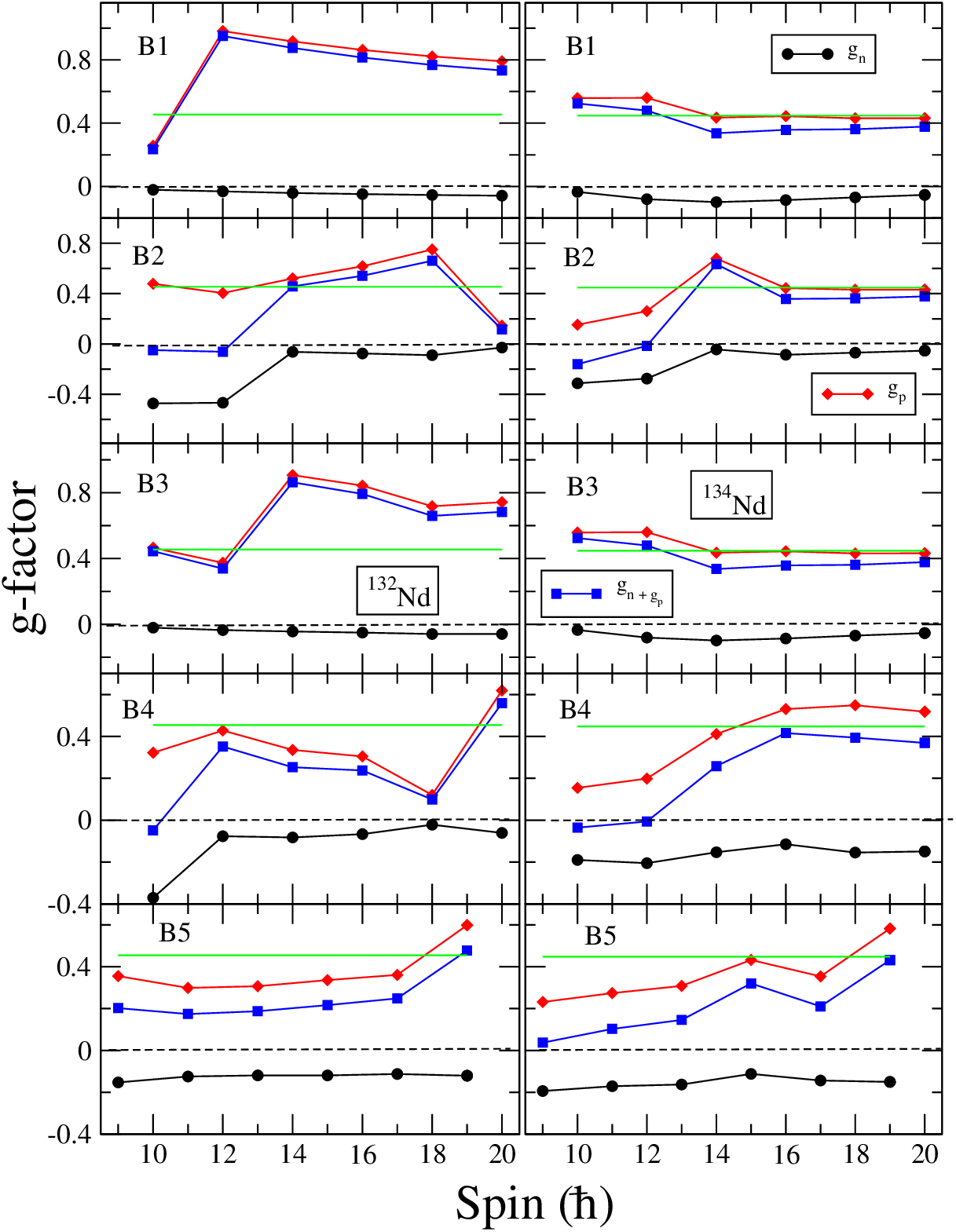}}
\caption{(Color online)  $^{132,134}$Nd g-factor with attenuated magnetic charges.  The green line shows $Z/A$.}
\label{fig12}
\end{figure}
\begin{figure}[htb]
 \centerline{\includegraphics[trim=0cm 0cm 0cm
0cm,width=0.9\textwidth,clip]{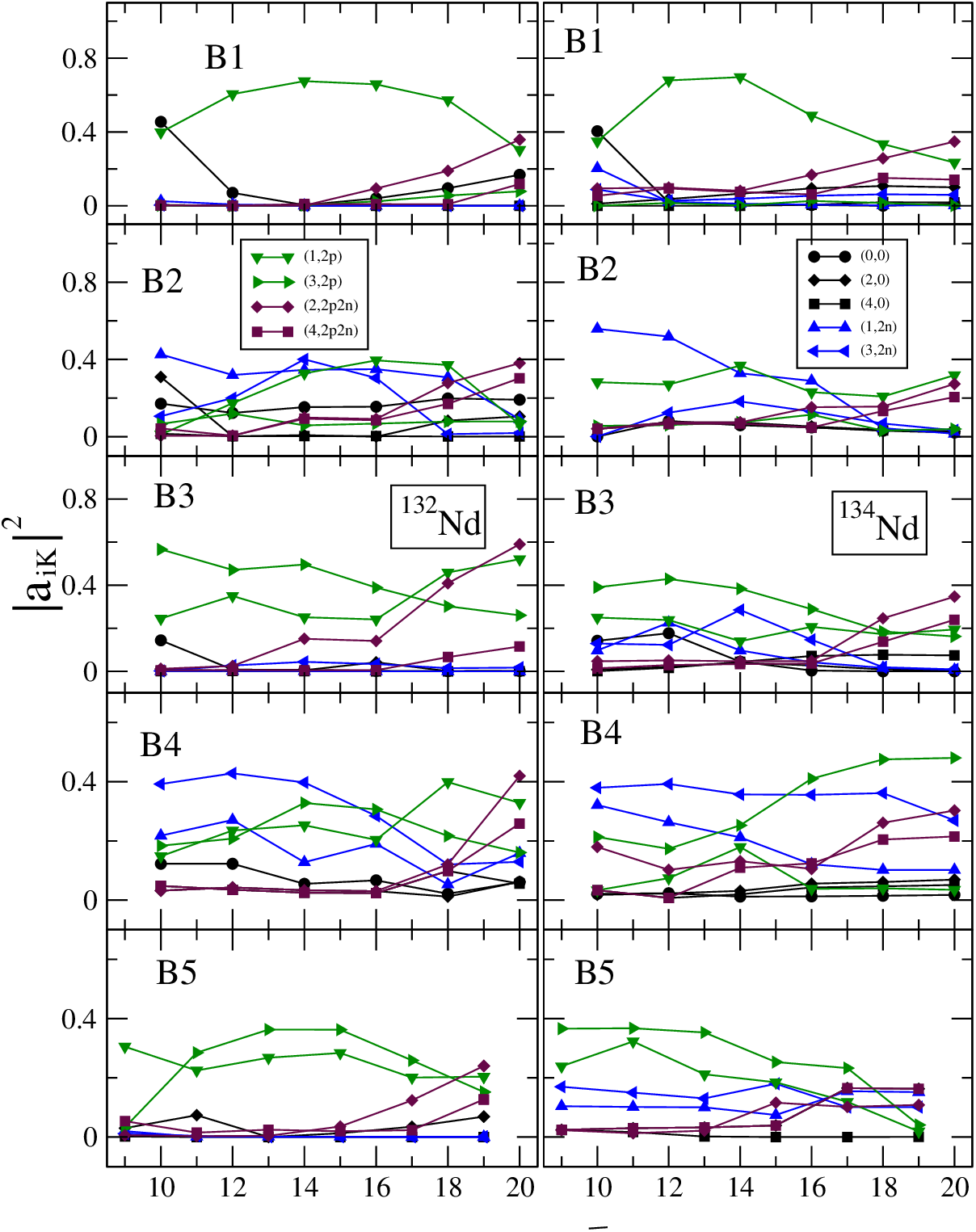}}
\caption{(Color online)   Probability of various projected K-configurations in the wavefunctions of the observed  bands for $^{132,134}$Nd. For clarity, only the lowest projected K-configurations in the wavefunctions of bands are shown and in the numerical calculations, projection has been performed from more than 
forty intrinsic states.}
\label{fig13}
\end{figure}

For Ce and Nd isotopes around A$\sim$130, forking of the ground-state
into s-bands has been observed \cite{wyss,Rj88} and in the present work
we have evaluated g-factors for these  band structures. It has
 been demonstrated in our earlier work \cite{JG09} that in some of these 
nuclei, $\gamma$-bands built on two-quasiparticle structure becomes
favoured in energy as for Ge-isotopes studied above. This lowering of
the $\gamma$-band was proposed to be the reason for observation of the
g-factors of the band heads of the two s-bands with the same sign
\cite{JG09}. However, g-factors were not evaluated in our
previous work and in the following g-factors shall be presented for
lowest s-band structures as has been done for Ge isotopes. 
\begin{figure}[htb]
 \centerline{\includegraphics[trim=0cm 0cm 0cm
0cm,width=0.9\textwidth,clip]{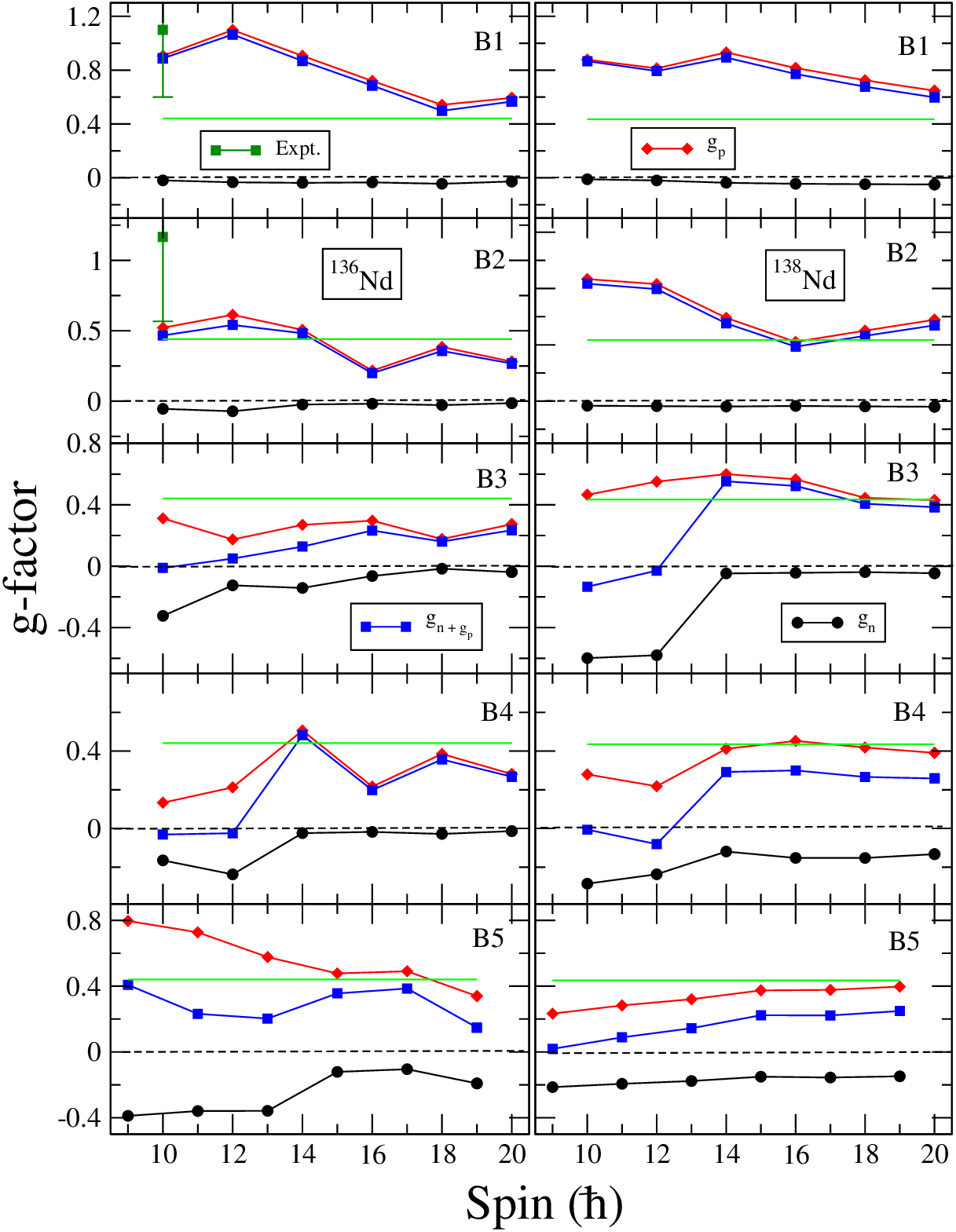}}
\caption{(Color online)  $^{136,138}$Nd g-factor with attenuated magnetic charges. The green line shows $Z/A$.}
\label{fig14}
\end{figure}
\begin{figure}[htb]
 \centerline{\includegraphics[trim=0cm 0cm 0cm
0cm,width=0.9\textwidth,clip]{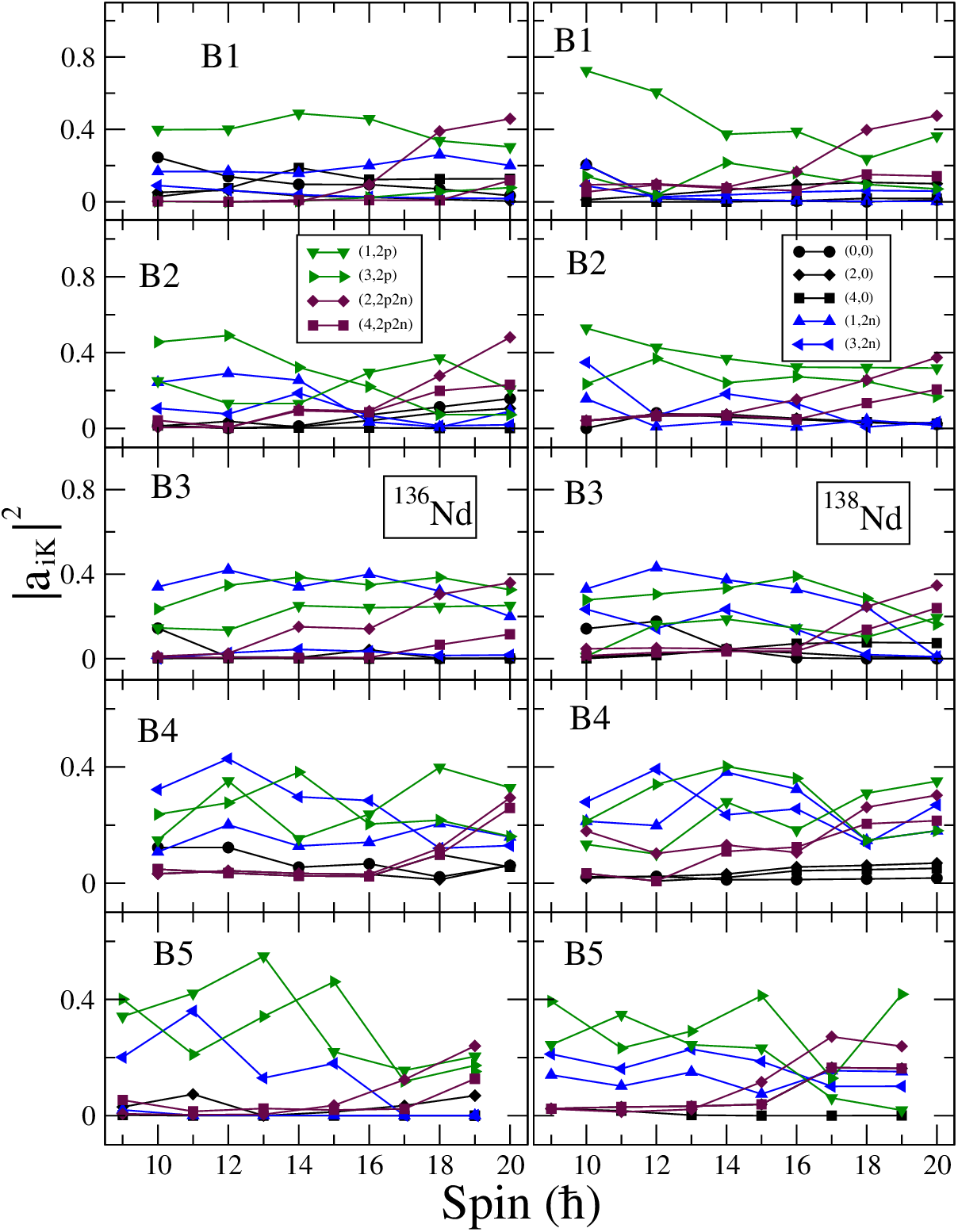}}
\caption{(Color online)   Probability of various projected K-configurations in the wavefunctions of the observed  bands for $^{136,138}$Nd. For clarity, only the lowest projected K-configurations in the wavefunctions of bands are shown and in the numerical calculations, projection has been performed from more than
forty intrinsic states.}
\label{fig15}
\end{figure}

For completeness, we shall first discuss the comparison between the observed 
and the TPSM calculated band structures of these nuclei before 
presenting the g-factors. The 
band structures for $^{128}$Ce, $^{130}$Ce, $^{132}$Ce and $^{134}$Ce  
are displayed in Figs.~\ref{fig4}, \ref{fig5}, \ref{fig6} and \ref{fig7}.  For all studied isotopes of Ce, multiple s-band structures are predicted and only for
  $^{132}$Ce
and $^{134}$Ce, two s-band
structures  are observed and it is noted from these figures that
observed energies are reproduced reasonably well by TPSM calculations.
 Also, TPSM calculated band structures for  $^{132}$Nd, $^{134}$Nd, $^{136}$Nd and $^{138}$Nd are presented in 
Figs.~\ref{fig4a}, \ref{fig5a}, \ref{fig6a} and \ref{fig7a}. The band structures show similiar behaviour as that of 
 Ce-isotopes and are  reproduced quite well by the calculations.

The calculated TPSM energies for all the isotopes studied
    in the present work are also displayed as numbers in Table 2 so
that other quantities of interest, for instance, the alignments and moments
of inertia, can be calculated and compared with other model predictions.

TPSM calculated g-factors for the $^{128}$Ce and $^{132}$Ce 
 are plotted
in Fig.~\ref{fig8} along with the  rigid value of Z/A.  The behaviour of the g-factors for $^{128}$Ce and $^{130}$Ce
is quite similar and can be easily understood from the corresponding wavefunction
plot of Fig.~ \ref{fig9}. It is noted from this figure that band B1 is
dominated by two-proton aligned configuration, $(1,2p)$ and the
corresponding total g-factor in Fig.~\ref{fig8} is 
positive. On the other hand, band B2 has dominant component of $(1,2n)$, which is 
a two-neutron aligned configuration, and the corresponding g-factor is
negative in Fig.~\ref{fig8}.  As already explained earlier, the
one-particle g-factor for proton is quite large as compared to the 
corresponding neutron g-factor, a small proton component in the
wavefuction makes the total g-factor skewed towards positive values.
Therefore, the total g-factor for band B2 in Fig.~\ref{fig8}, close to zero for spin values of I=10 and 12, indicates that this band, for these spin values, is
dominated by neutron configuration. For high-spin states, band B2 is
dominated by four-quasiparticle configurations and the total g-factor
tends to approach large positive values. Band B3 in Fig.~\ref{fig9} is
dominated by two-proton aligned states of $(1,2p)$ and $(3,2p)$
for $^{128}$Ce and  $^{130}$Ce, respectively and consequently g-factors in
Fig.~\ref{fig8} have large positive values. For bands B4 and B5, the
wavefunctions in Fig.~\ref{fig9} are highly mixed and the 
g-factors tend to be slightly positive.  However, for high-spin states,
the wavefunction is dominated by four-quasiparticle configurations
and the g-factors acquire large positive values.

TPSM calculated g-factors for $^{132}$Ce and $^{134}$Ce are
displayed in Fig.~\ref{fig10} along with the  rigid value
    of Z/A and the corresponding wavefunctions
are displayed in Fig.~\ref{fig11}. For band B1, the total g-factors
in Fig.~\ref{fig10} are negative for the low-spin states as the wavefunction
for this band is dominated by the neutron aligned configuration,
$(1,2n)$. There is a major difference  in the
wavefunctions of the band B2 for two nuclei. In $^{132}$Ce, band B2
is dominated by the aligned two-proton state, $(1,2p)$, whereas in 
$^{134}$Ce this band is dominated by $(3,2n)$, which is the
$\gamma$-band based on two-neutron aligned configuration. This
results into positive and negative g-factors of band B2 in two nuclei
for low-spin states in Fig.~\ref{fig10}.  Therefore, for $^{132}$Ce the two 
observed s-bands are expected to have opposite g-factors and for
$^{134}$Ce, both the s-bands should have negative g-factors as the two
bands have same intrinsic neutron configuration. As a matter
of the fact, this is confirmed by the experimental work performed in \cite{Ze82} and the observed  values are plotted in Fig.~\ref{fig10}.
Both the measured g-factors of the I=10 states of the two
s-bands being negative was an unresolved issue. It was explained
in our earlier publication and now backed up with numbers, that band B2 in $^{134}$Ce has the same intrinsic configuration
as that of band B1. The only difference is that band B2 is the
$\gamma$-band based on band B1.

In Fig.~\ref{fig11}, band B3 in $^{132}$Ce and $^{134}$Ce are dominated by
$(3,2p)$ and $(1,2p)$ configurations, respectively. Since both these are proton
configurations, the g-factors for this band are positive in
Fig.~\ref{fig10}. The wavefunctions for bands B4 and B5 in Fig.~\ref{fig11} are highly
mixed for two nuclei and the corresponding g-factors also reflect
this mixing. 

The g-factors for  $^{132}$Nd and $^{134}$Nd are shown in Fig.~\ref{fig12} 
along with the  rigid value of Z/A 
  and the corresponding wavefunctions are depicted in Fig.~\ref{fig13}. The 
g-factors for band B1 in both the nuclei are positive as the
wavefunction has the dominant proton aligned configuration,
$(1,2p)$. Band B2, on the other hand, is dominated by the neutron
aligned configuration, $(1,2n)$  for low-spin values and,
therefore, corresponding g-factors are negative for these spin
values. For high-spin states, four-quasiparticle configurations 
become important and consequently g-factors tend to acquire
positive values with increasing spin. Band B3 for both the nuclei
is mostly composed of $(3,2p)$ and, therefore, the g-factors
are positive for this band structure. As is evident from Fig.~\ref{fig12} 
that bands B4 and B5 are highly mixed with the consequence that 
g-factors are between neutron and proton maximal values.

The results for $^{136}$Nd and $^{138}$Nd are plotted in Figs.~\ref{fig14} 
and \ref{fig15}  along with the rigid value of Z/A. It is interesting to note from Fig.\ref{fig14} that g-factors for
the lowest two bands, B1 and B2, for both the nuclei are positive. The reason for this
is evident from the wavefunction decomposition in Fig.~\ref{fig15} with the
two bands dominated by $(1,2p)$ and $(3,2p)$, respectively. This
result is opposite to that obtained for $^{134}$Ce, where lowest two bands
have neutron configurations dominant. The experimental values available for the I=10 state in bands B1 and B2 for $^{136}$Nd are also positive and in good agreement with the calculated values.
Bands B3 and B4 in Fig.~\ref{fig14} have negative g-factors for low-spin
states as these bands have dominant two-neutron aligned
configurations. Band B5 is highly mixed with the g-factors 
in Fig.~\ref{fig14} depicting intermediate behaviour for some spin values.
For completeness, the attenuated g-factors for the ground-state bands
of all the nuclei studied in the present work are provided in Table 3.

\LTcapwidth=\textwidth
\begin{center}
\begin{longtable}{|p{0.05\textwidth}|p{0.11\textwidth}||p{0.05\textwidth}|p{0.11\textwidth}|p{0.11\textwidth}|p{0.11\textwidth}|p{0.11\textwidth}||p{0.05\textwidth}|p{0.11\textwidth}|}
  \caption{TPSM calculated energies for  g-band and excited band structures of $^{68,70}$Ge,  $^{128,130,132,134}$Ce, and $^{132,134,136,138}$Nd isotopes. }\\
  \hline
  


\hline
\endfirsthead

\multicolumn{7}{c}
{\tablename\ \thetable\ -- \textit{Continued from previous page}} \\
\hline


\hline
\endhead
\hline \multicolumn{9}{c}{\textit{}} \\
\endfoot
\hline
\endlastfoot
\multicolumn{9}{|c|}{$^{68}$Ge}\\ \hline
I($\hbar$) & g-band  &I($\hbar$) & B1  & B2 & B3& B4& I($\hbar$) & B5  \\\hline

0& 0.0   & 8    & 4.535  &5.098&5.691& 5.865&9&5.343\\
2&0.905    &10   & 5.497&6.323&6.689&6.932&11&6.397\\
4&2.170   &12   & 6.811&7.549&8.189&8.342&13&7.887\\
6&3.511    &14   & 8.404&8.634&9.428&9.743&15&9.442 \\
&    &16   & 9.881&10.231&10.655&11.254&17&11.239\\
&    &18   & 11.254&12.354&12.788&12.993&19&12.880\\
&    &20   & 12.986&14.086&15.016&15.121& &  \\
        \hline  
\multicolumn{9}{|c|}{$^{70}$Ge}\\ \hline      
 
0&0.0    & 8   &  3.854&4.708&4.898& 5.332&9&4.817\\
2&0.980    & 10  & 5.046&5.571&5.823& 6.383&11&6.034\\
4&1.996    & 12  & 6.583&6.738&7.238&7.538&13&7.284\\
6&3.199     & 14  &7.796&7.905&8.455&8.895&15& 8.592\\
&    & 16  &9.097&9.360&9.841&10.241&17& 9.897\\
&    & 18  &10.221&10.558&11.221&11.721&19& 11.408\\
&    & 20  &11.587&12.034&12.532&13.322&& \\
\hline 
\multicolumn{9}{|c|}{$^{128}$Ce}\\ \hline 
0&0.0    &  10 &2.515&2.871&3.407& 3.667&9&2.791\\
2& 0.234   &  12  &3.072 &3.547&3.851&4.229 &11&3.367\\
4& 0.566   &  14  &3.697 &4.330&4.621&4.883& 13&4.086\\
6& 1.122   &  16  &4.457 &5.169&5.389&5.566& 15&4.893\\
8& 1.806   &  18  &5.316 &5.975&6.182&6.334& 17& 5.735\\
&    &  20  &6.218 &6.615&6.977&7.097 &19&6.378\\
 \hline
\multicolumn{9}{|c|}{$^{130}$Ce}\\ \hline 
 
0&0.0  & 10 &  2.728 &2.975 &  3.500 &3.606&9 &2.775\\
2& 0.213    & 12 &  3.224 &3.707 &3.999 &4.329&11 &3.507\\
4&0.646     & 14 &  3.882 &4.436 & 4.675  &4.901  &13&4.236\\
6&1.258    & 16 &  4.693  &5.308&5.410 &5.573 &15& 5.008\\
8& 2.008    & 18 &  5.611  &5.997& 6.249 &6.438 &17&5.797\\
&    & 20 &  6.533  &6.842& 7.089 &7.284 & 19&6.642\\
\hline
\multicolumn{9}{|c|}{$^{132}$Ce}\\ \hline 
0&0.0   & 10 &2.988  & 3.408& 4.002&4.067& 9 &2.788\\
2& 0.250    & 12 &3.915  & 3.733& 5.006 &4.531& 11  & 3.715\\
4&  0.740   & 14 &4.774 & 4.417& 5.754 & 5.178&13 &4.574\\
6&  1.416  & 16 &5.718 & 5.191& 6.331 & 5.966&15 &5.418\\
8& 2.195      & 18 &6.582  & 6.019&7.193 &6.934&17 &6.383\\
&    & 20 & 7.521  & 6.877& 8.323 & 7.997&19  &7.321\\
 \hline
\multicolumn{9}{|c|}{$^{134}$Ce}\\ \hline 
  
0&  0.0   & 10 & 3.361  & 3.801& 4.378&4.398&9&3.517\\
2& 0.360    & 12 & 3.998  & 4.357&4.929 & 5.227&11& 4.029\\
4&0.983   & 14 &4.801 & 5.130& 5.588& 5.937&13&4.788\\
6&1.749    & 16 & 5.780  & 5.943& 6.215& 6.710&15&5.615\\
8& 2.567    & 18 & 6.566 & 6.945&7.250 &7.595&17& 6.505\\
&    & 20 &7.521& 8.023 & 8.428& 8.696&19&7.502\\
 \hline

\multicolumn{9}{|c|}{$^{132}$Nd}\\ \hline 
0&0.0    &  10   &2.414&2.743&3.404&3.597 &9& 2.543\\
2&0.156   &  12   &3.017&3.404&3.923&4.329&11&3.204\\
4&  0.504   &  14   &3.630&4.302&4.594&5.022&13&4.102\\
6& 1.018  &  16   &4.400 &5.204&5.529&5.641&15  & 5.004\\
8& 1.670    &  18   &5.312 &6.046&6.226&6.465 &17  &5.846 \\
&    &  20   &6.341 &6.831&7.111 &7.331&19 &6.531\\
  \hline
\multicolumn{9}{|c|}{$^{134}$Nd}\\ \hline 
0& 0.0  &10 &2.809 &2.971&3.595 & 3.833&9 & 2.714 \\
2& 0.216    &12 &3.349& 3.660&4.073&4.655&11 & 3.461\\
4&0.662   &14 &4.035 &4.433 &4.809&5.433&13  &4.233\\
6&1.294   &16 &4.827  &5.265 &5.673 &6.024&15  &5.065\\
8&2.047    &18 &5.661 &6.172 &6.618 &7.044&17  &6.017\\
&    &20 &6.526  &7.214 &7.694 &8.108&19 &7.014\\
\hline
\multicolumn{9}{|c|}{$^{136}$Nd}\\ \hline 
0& 0.0  & 10 &3.128& 3.403& 3.770& 4.295&9&3.203\\

2&0.356     & 12 &3.719&4.097 &4.341&4.946&11 &3.809 \\
4& 0.987  & 14 &4.596 & 4.956& 5.146&5.482&13&4.756\\
6&1.810    & 16 & 5.439& 5.693&5.977&6.423 &15 &5.593 \\
8&2.673     & 18 & 6.293&6.617&7.025&7.282& 17 &6.417\\
 &    &20 & 7.053&7.707& 8.100& 8.533&19 &7.307\\
 \hline
\multicolumn{9}{|c|}{$^{138}$Nd}\\ \hline 
0&0.0    & 10 &3.522 & 4.081&4.401&4.962 &9 &3.808\\
2& 0.489  & 12 &4.230&4.742&5.447 &5.831&11  &4.442\\
4& 1.225    & 14 &5.123&5.639&6.113&6.459 &13&5.139\\
6&2.140    & 16 &6.006&6.495&7.066&7.472&15  &6.095\\
8& 3.162   & 18 &7.065&7.453&8.298&8.512& 17&7.015\\
&    & 20 &8.337&8.514&9.497&9.951&19 & 8.014\\
 \hline
\end{longtable}
\end{center}


\begin{table}[htb]
\begin{center}

\caption{Calculated and measured g-factors for  ground-state bands of $^{68,70}$Ge,  $^{128,130,132,134}$Ce, and $^{132,134,136,138}$Nd isotopes.}
\vspace{0.9cm}
\begin{tabular}{|c|c|c|c|c|c|c|c|c|c|c|}
\hline
 I($\hbar$) &$^{68}$Ge  & $^{70}$Ge &$^{128}$Ce&$^{130}$Ce & $^{132}$Ce&$^{134}$Ce &$^{132}$Nd  & $^{134}$Nd&$^{136}$Nd&$^{138}$Nd\\ \hline
 2          
         & 0.02 &0.04&0.03&0.02 & 0.06 &0.04 &0.02 &0.05 &0.04 &0.07   \\ 
Expt.       &        &      &       &        &      &       &       &  0.6$^{+4}_{-4}$     &       &  \\     
 4        
            &-0.06 &-0.07&0.05&0.08  & 0.03 &0.05 &0.03 &0.05 &0.06&0.01    \\ 

 6       
            &-0.18 &-0.20 &0.02 &0.08  & -0.10 &-0.08 &0.06  &0.09   &0.03&0.03    \\ 

 8        
            &         &       & 0.15  &0.23&-0.17  &-0.08& 0.12 &0.17   &0.06 &0.27    \\ 
 \hline
\end{tabular}\label{gfground}
\end{center}
\end{table}

\section {Summary and Conclusions}

The purpose of the present work has been to perform a systematic analysis
of the g-factors of the excited band structures observed in selected isotopes
of Ge, Ce and Nd. These selected isotopes are predicted to depict forking of the
ground-state into several s-bands. In most of the nuclei, ground-state
band is crossed by quasiparticle aligned band, which in most of the
cases has a two-neutron character. However, for nuclei studied in
the present work, gound-state band is crossed by several band
structures simultaneously. For these nuclei, the Fermi surfaces of
neutrons and protons are similar and, therefore, both neutron and
proton aligned bands are expected to cross the ground-state band
simultaneously. In some of these nuclei, the excited bands depict
this expected structure. However, $^{68}$Ge and $^{134}$Ce  display two
s-bands with the same neutron structure and for $^{136}$Nd the bands have the same proton character.
The g-factors calculated using the TPSM wavefunctions reproduce the
neutron character for $^{68}$Ge and $^{134}$Ce, and proton structure for $^{136}$Nd.
 
It is evident from the TPSM
analysis that each triaxial intrinsic state has a $\gamma$- band built on it.
The triaxial ground- or vacuum-state is a superposition of K=0, 2,
4,... configurations and the angular-momentum projection 
of the K=0 state corresponds 
to the ground-state band. The projection of K=2 and 4 lead to the
$\gamma$- and $\gamma\gamma$-bands. In a similar manner, the 
two-quasiparticle aligned state is a superposition of K=1,3,5..... The 
angular-momentum projection of K=1 state leads to the normal s-band
and the projection with K=3 corresponds to the $\gamma$-band built
on the two-quasiparticle aligned state. In some nuclei, this
$\gamma$-band becomes favoured as compared to other configurations
and crosses the ground-state band. In $^{68}$Ge and $^{134}$Ce, the
lowest two aligned bands that cross the ground-state band are the
two-neutron aligned band and the $\gamma$-band built on this state.
For $^{136}$Nd the lowest two bands have proton character. Since these bands have the same intrinsic configuration, g-factors
of the two band structures are also expected to be similar. Therefore,
TPSM analysis provides a natural explanation for the observation of
neutron g-factors for the bandheads of two lowest aligned band
structures in $^{68}$Ge and $^{134}$Ce and proton for $^{136}$Nd.

Further, it has been predicted that observed two s-bands in $^{70}$Ge 
should also have negative g-factors. For the case of $^{138}$Nd, the lowest two s-bands are predicted
to have positive g-factors. It would be quite interesting to perform the
g-factor measurements of these nuclei to verify the predictions of the
present model  analysis. It would be also interesting to explore other 
nuclei having $\gamma$-bands based on excited quasiparticle configurations.

\end{document}